\documentclass[submission, Phys]{SciPost}

\usepackage{braket}
\usepackage{amsmath,amssymb}
\usepackage{graphicx}
\usepackage{floatrow}

\begin{document}

\begin{center}{\Large \textbf{
Quantum quench dynamics in the transverse-field Ising model:  
A numerical expansion in linked rectangular clusters}}\end{center}

\begin{center}
Jonas Richter\textsuperscript{*},
Tjark Heitmann,
Robin Steinigeweg
\end{center}

\begin{center}
 Department of Physics, University of Osnabr\"uck, D-49069 Osnabr\"uck, Germany
\\

* jonasrichter@uos.de
\end{center}

\begin{center}
\today
\end{center}


\section*{Abstract}
{\bf

We study quantum quenches in the transverse-field 
Ising model defined on different lattice 
geometries such as chains, two- and three-leg ladders, 
and two-dimensional square lattices. 
Starting from fully polarized initial 
states, we consider the dynamics of 
the transverse and the longitudinal 
magnetization for quenches to weak, strong, and critical 
values of the transverse field. To this end, we rely on an efficient 
combination of numerical linked cluster 
expansions (NLCEs) and a forward propagation of pure states in 
real time.
As a main result, we demonstrate that NLCEs comprising solely 
rectangular clusters 
provide a promising approach to study 
the real-time dynamics of two-dimensional quantum many-body systems directly in 
the thermodynamic limit. By comparing to existing data from the 
literature, we unveil that NLCEs yield converged results on time scales 
which are competitive to other state-of-the-art numerical methods. 

}

\vspace{10pt}
\noindent\rule{\textwidth}{1pt}
\tableofcontents\thispagestyle{fancy}
\noindent\rule{\textwidth}{1pt}
\vspace{10pt}

\section{Introduction}
\label{Sec::Intro}

Understanding the dynamics of isolated quantum many-body systems out of 
equilibrium is an active area of research of modern theoretical and 
experimental physics \cite{Polkovnikov_2011,Eisert_2015,Bertini2020}. A 
popular nonequilibrium protocol in this context 
is a 
so-called quantum quench \cite{Mitra_2018}. In such quench protocols, the 
system's 
Hamiltonian 
${\cal H}$ depends on some parameter $\lambda$, and the system is 
prepared 
in an eigenstate $\ket{\psi(0)}$ of ${\cal H}$, e.g., the groundstate, for an 
initial value 
$\lambda_i$. Next, the value of $\lambda$ is suddenly changed, $\lambda_i \to 
\lambda_f$, such that $\ket{\psi(0)}$ is no eigenstate of ${\cal H}(\lambda_f)$, 
and the system exhibits nontrivial dynamics. 
For an isolated quantum system undergoing unitary time evolution, it is then 
intriguing to study if and in which way the system relaxes back to 
equilibrium. Central questions are, for instance, how the (short- 
or long-time) dynamics can be described in terms of ``universal'' principles 
\cite{Polkovnikov_2011,Dziarmaga_2010,Reimann_2016,Erne_2018,Pr_fer_2018,
Richter_2019_3,
Richter_2019_4,Dupont_2020}, what are the relevant time scales of relaxation 
\cite{Garc_a_Pintos_2017,Wilming_2018,Alhambra_2020}, and whether 
or not the long-time values of physical observables agree 
with the prediction of, e.g., a microcanonical or canonical 
ensemble (i.e.\ thermalization) \cite{D_Alessio_2016, 
Borgonovi_2016, Gogolin_2016}.

One possible mechanism to explain the emergence of thermalization in isolated 
quantum systems is given by the eigenstate thermalization hypothesis (ETH)
\cite{Deutsch_1991, Srednicki_1994, Rigol_2008}. While the validity of the ETH 
has been numerically tested for a variety of models and observables (see, 
e.g., \cite{Santos_2010, Steinigeweg_2013, Beugeling_2014, Mondaini_2016, 
Mondaini_2017, Kim_2014, Jansen_2019}), there are 
also classes of systems which violate the ETH and fail to  
thermalize. One such class is given by integrable models, where the  
extensive number of conservation laws prevents the applicability of standard 
statistical ensembles \cite{Essler_2016}. 
Instead, 
it has been proposed that integrable models equilibrate towards a 
generalized Gibbs ensemble (GGE), which maximizes the entropy with respect to 
the 
conserved charges \cite{Jaynes_1957,Rigol_2007,Vidmar_2016}. In addition, it is 
now widely believed that some strongly 
disordered systems can undergo a transition to a many-body localized 
(MBL) 
phase, where the ETH is violated as well \cite{Nandkishore_2015, Abanin_2019}.
Moreover, there 
has been plenty 
of interest 
recently in models which are, in a sense, intermediate cases between 
``fully ETH'' and ``fully MBL''. This includes, e.g., models featuring 
``quantum scars'' where rare ETH-violating 
states are embedded in an otherwise thermal spectrum 
\cite{Shiraishi_2017, Turner_2018, Schecter_2019, Iadecola_2019,
Lee_2020}, as 
well as models 
which exhibit a strong fragmentation of 
the Hilbert space due to additional contraints \cite{Khemani_2019, Sala_2020}.  

From a numerical point of view, studying the nonequilibrium dynamics of
isolated quantum many-body systems is a challenging task. This is not 
least caused by the fact that for an interacting quantum system, the 
Hilbert space grows exponentially in the number of constituents. 
Nevertheless, thanks to the continuous increase of computational resources 
and the development of sophisticated numerical 
methods including, e.g., dynamical mean field theory \cite{Aoki_2014}, Krylov 
subspace techniques \cite{Nauts_1983,Long_2003}, dynamical quantum typicality 
\cite{Heitmann_2020}, or classical representations in phase space 
\cite{Wurtz_2018}, significant progress has been made. Especially for 
one-dimensional systems, the time-dependent density-matrix 
renormalization group, as well as related methods based on 
matrix-product states (MPS), provide a powerful tool to study dynamical 
properties for system sizes practically in the thermodynamic limit 
\cite{Schollw_ck_2011, Paeckel_2019}. 
However, since these methods rely on an efficient compression of moderately  
entangled wavefunctions, the reachable time scales in simulations are 
eventually limited due to the inevitable buildup of entanglement during 
the unitary time evolution. 

The growth of entanglement becomes even more severe in spatial dimensions 
larger than one. Despite recent advances involving MPS-based or tensor-network 
algorithms \cite{Zaletel_2015, James_2015,  Hashizume_2018, Hubig_2019, 
Czarnik_2019, Kloss_2020}, as 
well as the advent of innovative 
machine-learning approaches \cite{Carleo_2017, Schmitt_2018, Schmitt_2019}, 
the time scales numerically attainable for
two-dimensional quantum many-body systems are still comparatively short. While 
the dynamics of such two-dimensional systems can nowadays 
be accessed in experiments with quantum simulators 
\cite{Choi_2016, Guardado_Sanchez_2018,Lienhard_2018}, 
the development 
of efficient numerical techniques is paramount. On the one 
hand, unbiased numerical simulations are important to confirm the accuracy of 
the 
experimental results. On the other hand, numerical simulations can also serve 
as an orientation for experiments to explore certain models or parameter 
regimes in more detail.

In this paper, we scrutinize the nonequilibrium dynamics for quantum quenches in 
the Ising model with transverse 
magnetic 
field. While this model is 
exactly solvable in the case of a chain and has been studied in numerous 
instances, our main focus is 
on
nonintegrable geometries such as two- and 
three-leg ladders and, in particular, two-dimensional square lattices.   
To this end, we rely on an efficient combination of numerical
linked cluster expansions (NLCEs) and the iterative forward propagation of 
pure states 
in real time via Chebyshev polynomials.   
Depending on 
the model geometry, the initial state, and the strength of the quench,
the nonequilibrium dynamics is found to display a variety of different 
behaviors 
ranging 
from rapid equilibration, over slower 
monotonous relaxation, to persistent (weakly damped) oscillations. 
Most importantly, from a 
methodological point of view, we demonstrate that NLCEs comprising 
solely 
rectangular clusters 
provide 
a promising 
approach to study the real-time dynamics of two-dimensional quantum many-body 
systems directly in the thermodynamic limit. By comparing to existing data 
from the literature, we unveil that NLCEs yield converged results 
on time scales which are competitive to other 
state-of-the-art 
numerical methods. 

This paper is structured as follows. In Sec.\ \ref{Sec::Model}, we introduce 
the models, observables, and quench protocols which are studied. In 
Sec.\ \ref{Sec::Numerics}, we then discuss the employed 
numerical methods, while our results are presented in Sec.\ \ref{Sec::Results}.
We summarize and conclude in Sec.\ \ref{Sec::Conclusion}.

\section{Models, observables, and quench protocols}\label{Sec::Model}

We study the Ising model with ferromagnetic nearest-neighbor interactions 
and transverse magnetic field, described by the Hamiltonian
\begin{equation}\label{Eq::Hamiltonian}
 {\cal H} = -J \left(\sum_{\langle \ell, m \rangle} \sigma_\ell^z \sigma_m^z +
g \sum_{\ell = 1}^L 
\sigma_\ell^x\right)\ ,  
\end{equation}
where the first sum on the right hand side runs over all pairs of nearest 
neighbors $\ell$ and $m$, $L$ is the total number of sites, $J>0$ sets the 
energy 
scale, $g>0$ denotes the strength of the transverse field, and 
$\sigma_\ell^{x,z}$ 
are Pauli matrices at site $\ell$. Note that the Hamiltonian 
\eqref{Eq::Hamiltonian} is symmetric under the global spin-flip operation 
$\sigma_{\ell}^z \to -\sigma_{\ell}^z$.

In this paper, the transverse-field Ising 
model \eqref{Eq::Hamiltonian} is considered for different lattice 
geometries such as chains $(L = L_x)$, two- and three-leg ladders $(L =  
L_x \times 2, L = L_x \times 3)$, and two-dimensional 
square lattices $(L = L_x \times L_y)$. While we generally intend to 
obtain results in the thermodynamic limit $L \to \infty$ (see Sec.\ 
\ref{Sec::Numerics_NLCE} for our numerical approach), we consider finite 
system sizes as well. 
In the case $L < \infty$, one has to distinguish between open boundary 
conditions (OBC) and 
periodic boundary conditions (PBC), where for chains and ladders the latter 
only 
applies in the $x$ direction.  

On the one hand, in the case of a chain, ${\cal H}$ is a paradigmatic example 
of an integrable model and can be solved exactly by subsequent Jordan-Wigner, 
Fourier, and Bogolioubov transforms \cite{Pfeuty_1970}, see also 
Appendix \ref{App::1d}. 
For $g < 1$, ${\cal H}$ is in a ferromagnetic phase with a two-fold degenerate 
groundstate. At the critical point $g = 1$, ${\cal H}$ undergoes a quantum 
phase transition towards a paramagnetic phase with unique groundstate for $g > 
1$. On the other hand, for a two-dimensional square lattice, ${\cal H}$ is 
nonintegrable \cite{Mondaini_2016, Mondaini_2017,Bla__2016}, and the quantum 
phase 
transition between an ordered 
phase and an unordered phase occurs at the larger transverse field $g = g_{c} 
\approx 3.044$ \cite{Bl_te_2002}. 
For intermediate cases, such as multi-leg 
ladders on a cylinder geometry, the value of $g_c$ can vary since 
these cases are quasi-one-dimensional \cite{Hashizume_2018}.

In this paper, we consider quench protocols starting from fully polarized 
initial states 
$\ket{\psi(0)}$. Namely, we either study quenches starting 
from  
$\ket{\psi(0)} = \ket{\uparrow}$, 
\begin{equation}
 \ket{\uparrow} = \ket{\uparrow \uparrow \cdots \uparrow}\ , 
\end{equation}
where all spins are initially aligned along the $z$ axis, or quenches 
starting from the state $\ket{\psi(0)} = \ket{\rightarrow}$,  
\begin{align}
 \ket{\rightarrow} = \ket{\rightarrow \rightarrow \cdots \rightarrow}\ ,
\end{align}
where all spins point in the $x$ direction. Note that written in the common 
eigenbasis of the local $\sigma_\ell^z$, $\ket{\rightarrow}$ is a uniform 
superposition of all $2^L$ basis states. Moreover, while the state 
$\ket{\uparrow}$ is 
an eigenstate of ${\cal H}$ for vanishing field $g = 0$, the state 
$\ket{\rightarrow}$ is the groundstate of ${\cal H}$ 
for $g \to \infty$. 
Given the states $\ket{\uparrow}$ and $\ket{\rightarrow}$, we study the 
nonequilibrium dynamics 
resulting from 
quantum quenches to weak $(g < g_c)$, strong $(g > g_c)$, or critical values 
$(g 
= g_c)$ of the transverse field, i.e., depending on the initial state
these are quenches either within the same equilibrium phase, or to or 
across the critical point.  

Due to the quench, the fully polarized states $\ket{\uparrow}$ and 
$\ket{\rightarrow}$ are no eigenstates 
of ${\cal H}$ anymore and evolve unitarily in time $(\hbar = 1)$,
\begin{equation}\label{Eq::TimeEvo}
 \ket{\psi(t)} = e^{-i{\cal H}t} \ket{\psi(0)}\ . 
\end{equation}
Consequently, the expectation values of 
observables acquire a dependence on time as well. In particular, we 
here consider the 
dynamics of the transverse and the longitudinal magnetization,
\begin{equation}
 \langle X(t) \rangle = \frac{1}{L}\sum_{\ell=1}^L 
 \bra{\psi(t)} \sigma_\ell^{x} \ket{\psi(t)}\ ,\quad \langle Z(t) \rangle = 
\frac{1}{L}\sum_{\ell=1}^L 
 \bra{\psi(t)} \sigma_\ell^{z} \ket{\psi(t)}\ .
\end{equation}

\section{Numerical approach}\label{Sec::Numerics}

We now discuss the numerical methods which are employed in this paper. 
Throughout this section, we exemplarily focus on the transverse magnetization 
$\langle X(t) \rangle$. The calculations for $\langle Z(t) \rangle$ are carried 
out analogously. 

\subsection{Numerical linked cluster expansion}\label{Sec::Numerics_NLCE}

Numerical linked cluster expansions provide a means to access the properties of 
quantum many-body systems directly in the thermodynamic limit. Originally 
introduced to study thermodynamic quantities \cite{Rigol_2006, Rigol_2007_2} 
(see also \cite{Ixert_2015,Bhattaram_2019,Schaefer_2020}), 
NLCEs have more recently 
been 
employed to obtain entanglement entropies \cite{Kallin_2013}, to calculate 
steady-state 
properties in driven-dissipative systems \cite{Biella_2018}, to study quantum 
quenches 
with mixed or 
pure initial states 
\cite{Rigol_2014,Wouters_2014,White_2017,Mallayya_2017,Mallayya_2018,
Guardado_Sanchez_2018}, as well as to simulate time-dependent 
equilibrium correlation functions \cite{Richter_2019, Richter_2019_2}. 

The main idea of NLCEs is that the per-site value of an extensive quantity in 
the thermodynamic limit can be obtained as a sum over contributions from all 
linked clusters which can be embedded on the lattice \cite{Tang_2013},
\begin{equation}\label{Eq::NLCE}
 \lim_{L\to\infty} \langle X(t) \rangle = \sum_c {\cal L}_c W_c(t)\ , 
\end{equation}
where the sum runs over all connected clusters $c$ with multiplicities ${\cal 
L}_c$ 
and weights $W_c(t)$. Specifically, ${\cal L}_c$ is the number 
of ways (normalized by the size of the lattice) a cluster $c$ can be 
embedded on the lattice [see also the discussion 
around Eq.\ \eqref{Eq::Multi} below].
Moreover, the notion of a \textit{connected} cluster 
refers to a 
finite 
number of lattice sites, where every site of the cluster has to be 
directly connected to 
at least one other cluster site by terms of the underlying Hamiltonian.
Given a two-dimensional square lattice and the 
nearest-neighbor Hamiltonian in Eq.\ \eqref{Eq::Hamiltonian}, for instance, 
the lattice sites 
$(i,j)$ and $(i,j+1)$ form a connected cluster of size two. In contrast, the 
sites $(i,j)$ and $(i+1,j+1)$ do not form a connected cluster as ${\cal H}$ 
does not contain terms along the diagonal. However, in combination, the sites 
$(i,j)$, $(i,j+1)$, and $(i+1,j+1)$ would be a connected cluster of size 
three.

Given a cluster $c$, its weight $W_c(t)$ is 
obtained by an 
inclusion-exclusion principle. That is, the quantity of interest (here 
the dynamics of the magnetization $X$) is evaluated on the cluster $c$ (with 
OBC) and, subsequently, the weights $W_s(t)$ of all subclusters $s$ of $c$ have 
to be subtracted \cite{Tang_2013},  
\begin{equation}\label{Eq::Inclu}
 W_c(t) = \langle X(t) \rangle_{(c)} - \sum_{s\subset c} W_s(t)\ . 
\end{equation}

While NLCEs yield results in the thermodynamic limit (such that a finite-size 
scaling becomes unnecessary), it is instead crucial to check the convergence of 
the series. 
To this end, the sum in Eq.\ \eqref{Eq::NLCE} is usually organized in terms of 
expansion 
orders \cite{Tang_2013}. For instance, one could group together all clusters 
which comprise a certain number of lattice sites. 
Then, an expansion up to order $C$ refers to the fact that all  
clusters with up to $C$ lattice sites are considered in Eq.\ 
\eqref{Eq::NLCE}. Moreover, the NLCE is said to be converged if the outcome 
of Eq.\ \eqref{Eq::NLCE} does not depend on the value of $C$.   
\begin{figure}[tb]
 \centering
 \includegraphics[width=1\textwidth]{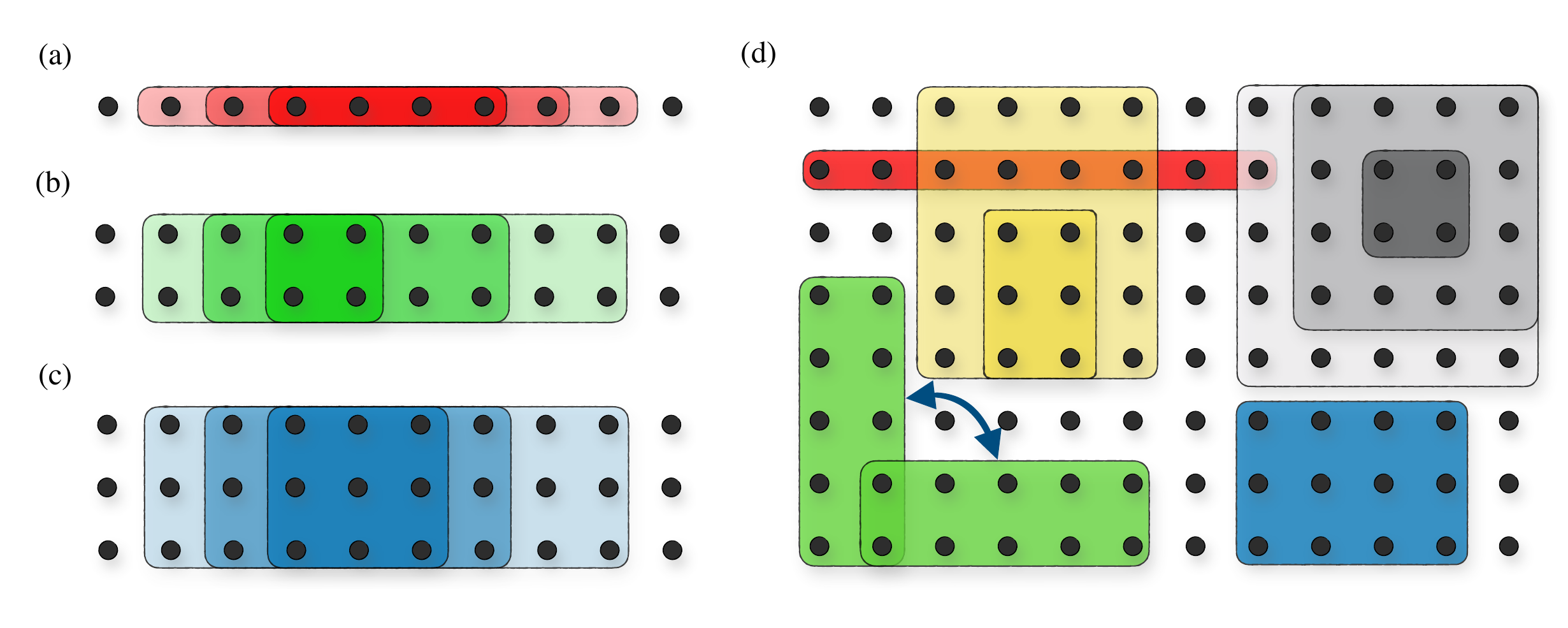}
 \caption{Examples of clusters which are used in the 
NLCE. (a) For a chain geometry, all clusters and subclusters are chains. 
(b) and (c) In case of a ladder 
geometry, we only consider clusters and subclusters which are ladders as 
well. 
(c) For the two-dimensional square lattice, we restrict 
ourselves to clusters with a rectangular shape. Given the 
Hamiltonian ${\cal H}$ in Eq.\ \eqref{Eq::Hamiltonian}, we note that a cluster 
$c=(x,y)$ with 
$x > y$ is equivalent to its $90^\circ$-rotated counterpart $c^\prime = 
(y,x)$. To speed up the simulations, we therefore only need to consider 
clusters $c$ with $x \geq y$, where square-shaped clusters with $x = y$ enter 
Eq.\ \eqref{Eq::NLCE} once, while rectangular clusters with $x > y$ enter 
the expansion 
twice.}
 \label{Fig1}
\end{figure}

At this point, it is important to note that in actual simulations, the maximum 
order $C$ that can be reached 
is limited by two factors: (i) the exponential growth of the Hilbert-space 
dimension with increasing cluster size, and (ii) the necessity to identify the 
(possibly very 
large number of) 
distinct clusters and to calculate their weights.
Since a larger expansion order typically leads to a convergence of Eq.\ 
\eqref{Eq::NLCE} up to longer times \cite{Richter_2019} (or down to lower 
temperatures for 
thermodynamic quantities \cite{Bhattaram_2019}), it is desirable to include 
clusters as large 
as possible. 
In this paper, we therefore aim to mitigate the limitations (i) and (ii) by two 
complementary approaches. First, instead of using full exact 
diagonalization to evaluate $\langle X(t) \rangle_{(c)}$, we here employ an 
efficient forward propagation of pure states (see Sec.\ 
\ref{Sec::Numerics_PSP}), which is feasible for significantly larger 
Hilbert-space dimensions.
Secondly, in order to reduce the enormous combinatorial costs to generate (and 
evaluate) all clusters with a given number of sites, we rely on the fact 
that the 
sum in Eq.\ \eqref{Eq::NLCE} can also converge for different types of 
expansions, as long as clusters and 
subclusters can be defined in a self-consistent manner \cite{Tang_2013}.
In this paper,  
we specifically
restrict ourselves to only those clusters which have a rectangular shape.
This restriction is particularly appealing as the number of 
distinct clusters is 
significantly reduced and the
calculation of the weights $W_c(t)$ becomes rather simple since all subclusters 
are rectangles as well, see Fig.\ \ref{Fig1}. 
Furthermore, the rectangle expansion has been succesfully used before to 
obtain entanglement entropies \cite{Kallin_2013}, and it also appears to be a 
promising candidate to study dynamical properties as it involves 
clusters with many different length scales. 
In this context, let us note that other restricted expansions 
for the two-dimensional square lattice, e.g., clusters consisting of 
corner-sharing $2\times 2$ squares, have proven to be a good choice to extract 
thermodynamic quantities \cite{Bhattaram_2019}. In this paper, however, we 
focus on rectangular clusters as a first case study.  

Given a rectangular cluster $c = (x,y)$ of width $x$ and 
height $y$, the inclusion-exclusion principle from 
Eq.\ \eqref{Eq::Inclu} to obtain the weight $W_{(x,y)}(t)$ takes on the form 
\cite{Dusuel_2010}
\begin{equation}\label{Eq::WeightRec}
 W_{(x,y)}(t) = \langle X(t) \rangle_{(x,y)} - 
\underset{x^\prime y^\prime < xy}{\sum_{x^\prime=1}^x 
\sum_{y^\prime 
= 1}^y} (x - x^\prime + 1) (y-y^\prime + 1) W_{(x^\prime,y^\prime)}\ ,
\end{equation}
where the sum runs over all rectangular subclusters.
Next, in order to carry out the expansion \eqref{Eq::NLCE}, the 
multiplicity ${\cal L}_{c}$ is required. Given a two-dimensional square lattice 
of size $L_x\times L_y$ with OBC, the number of ways per lattice site a 
rectangular cluster $c=(x,y)$ (with finite $x$ and $y$) can be embedded 
on the lattice follows as,
\begin{equation}\label{Eq::Multi}
 {\cal L}_c = \frac{(L_x-x+1)(L_y-y+1)}{L_xL_y}\ , 
\end{equation}
as there are $(L_x-x+1)$ possible translations in the $x$ 
direction and $(L_y-y+1)$ in the $y$ direction. Thus, if one is interested in 
the properties of the lattice in the thermodynamic limit, $L_x,L_y \to \infty$, 
one finds,
\begin{equation}
 {\cal L}_c = 1\ .
\end{equation}

Furthermore, in order to speed up the simulations, 
it is useful to take into account that the Hamiltonian ${\cal H}$ in Eq.\ 
\eqref{Eq::Hamiltonian} is invariant under rotations in the sense that a 
rectangular cluster $c = (x,y)$ with $x \geq y$ yields the exact same weight 
$W_c(t)$ as the cluster 
$c^\prime = (y,x)$, i.e., $c$ rotated by $90$ degrees. Thus, in practice, we 
only need to consider clusters with $x \geq y$, where square-shaped cluster 
with $x = y$ enter Eq.\ \eqref{Eq::NLCE} once, while rectangular clusters with 
$x > y$ then enter the expansion twice.

Let us add some comments on the NLCE for chains 
and ladders. First, we note that in the case of chains, all clusters are just 
chains as well, see Fig.\ \ref{Fig1}~(a). In this case, the expansion in  Eq.\ 
\eqref{Eq::NLCE} reduces to 
a single 
difference between $\langle X(t)\rangle_{(c)}$ evaluated on the largest and the 
second-largest cluster \cite{Richter_2019}.
Secondly, while for two-leg (or three-leg) ladders, 
rectangular clusters can in principle have a height $y = 1,2$ (or $y = 1,2,3$)
with different lengths $x$, we here restrict ourselves even further to those 
clusters which are two-leg or three-leg ladders as well, see Figs.\ 
\ref{Fig1}~(b) and (c). In this case, the expansion \eqref{Eq::NLCE} again 
reduces to a single difference between $\langle X(t)\rangle_{(c)}$ evaluated on 
the largest and the 
second-largest cluster. Despite this simplicity, however, we find that this 
type of expansion for ladders in 
practice yields convincing convergence times. 

\subsection{Pure-state propagation}\label{Sec::Numerics_PSP}

Evaluating the unitary time evolution of the initial states 
$\ket{\psi(0)}$ according to Eq.\ \eqref{Eq::TimeEvo} in principle requires the 
full
exact 
diagonalization (ED) of the 
Hamiltonian ${\cal H}$. In order to access system (and cluster) sizes beyond 
the range of full
ED, we here subdivide the evolution up to time $t$ into a product of discrete 
time steps,  
\begin{equation}\label{Eq::TimeEvo}
\ket{\psi(t)} = e^{-i{\cal H}t}\ket{\psi(0)} =  \left(e^{-i{\cal H}\delta 
t}\right)^Q \ket{\psi(0)}\ ,  
\end{equation}
where $\delta t = t/Q$. If the time step $\delta t$ is chosen 
sufficiently small, then there exist various approaches to accurately 
approximate the 
action of the exponential $\exp(-i{\cal H}\delta t)$ such as, e.g., Trotter 
decompositions \cite{de_Vries_1993}, Krylov subspace techniques 
\cite{Nauts_1983}, or 
Runge-Kutta 
schemes \cite{Elsayed_2013, Steinigeweg_2014}. In this paper, we rely on an 
expansion of the time-evolution 
operator 
in terms of Chebyshev polynomials, for a comprehensive overview see 
\cite{Tal_Ezer_1984, Dobrovitski_2003, Wei_e_2006, Fehske_2009}. 
Let us emphasize that the evaluation of Eq.\ 
\eqref{Eq::TimeEvo} to a high precision is crucial for the convergence of 
the NLCE. Even relatively small numerical errors for the contribution of each 
individual cluster could eventually spoil the convergence of the series when 
combined according to Eq.\ \eqref{Eq::NLCE}. In this context, the Chebyshev 
polynomial expansion is known to yield very accurate results for a given step 
size $\delta t$. In contrast, we have checked that reaching the same level of 
accuracy by means of a fourth-order Runge-Kutta scheme requires a significantly 
smaller $\delta t$ which, in turn, increases the overall runtime of 
the simulation.

Since the Chebyshev polynomials are defined on the interval $[-1,1]$, the 
spectrum of the original Hamiltonian ${\cal H}$ has to be rescaled 
\cite{Fehske_2009},  
\begin{equation}
 \widetilde{{\cal H}} = \frac{{\cal H} - b}{a}\ , 
\end{equation}
where $a$ and $b$ are suitably chosen parameters. In practice, we use the fact 
that the (absolute of the) extremal eigenvalue of ${\cal H}$ can be bounded 
from above according to \cite{Dobrovitski_2003}  
\begin{equation}
\max(|E_\text{min}|,|E_\text{max}|) \leq J \left(N_{\langle 
\ell,m\rangle} + gL\right) = {\cal E}\ ,
\end{equation}
where $E_\text{max}$ ($E_\text{min}$) is the largest (smallest) eigenvalue of 
${\cal H}$, and $N_{\langle \ell,m\rangle}$ denotes the number of 
nearest-neighbor pairs 
$\langle \ell,m \rangle$, i.e., the number of bonds of the lattice. 
By choosing 
$a \geq {\cal E}$, it is guaranteed that the spectrum of ${\widetilde{\cal 
H}}$ lies within $[-1,1]$. As a consequence, we can set $b = 0$. Note 
that while this choice of $a$ and $b$ is not necessarily 
optimal, it proves to be sufficient \cite{Dobrovitski_2003} (see also Appendix 
\ref{App::AccuPSP}). 

Within the Chebyshev-polynomial formalism, the time evolution of a state 
$\ket{\psi(t)}$ can then be approximated as an 
expansion up to order $M$ \cite{Fehske_2009}, 
\begin{equation}\label{Eq::Cheb1}
 \ket{\psi(t+\delta t)} \approx c_0\ket{v_0} + \sum_{k=1}^M 2c_k\ket{v_k}\ , 
\end{equation}
where the expansion coefficients $c_0, c_1, \dots, c_M$,  
are given by
\begin{equation}\label{Eq::ChebCoeff}
 c_k = (-i)^k {\cal J}_k(a\delta t)\ , 
\end{equation}
with ${\cal J}_k(a\delta t)$ being the $k$-th order Bessel function of the 
first 
kind evaluated at $a\delta t$. 
[Note that the notation in Eqs.\ \eqref{Eq::Cheb1} and \eqref{Eq::ChebCoeff} 
assumes $b = 0$.] Moreover, the vectors 
$\ket{v_k}$ 
are recursively generated according to 
\begin{equation}\label{Eq::Cheb2}
 \ket{v_{k+1}} = 2\widetilde{{\cal H}}\ket{v_k} - \ket{v_{k-1}}\ ,\ \quad k 
\geq 1\ , 
\end{equation}
with $\ket{v_1} = \widetilde{{\cal H}}\ket{v_0}$ and $\ket{v_0} = 
\ket{\psi(t)}$. Given a time step $\delta t$ (and the parameter $a$), 
the expansion order $M$ has to be chosen large enough to ensure negligible 
numerical errors. In this paper, we typically have $\delta t J = 0.02$ 
and $M = 15$, which turns out to yield very accurate results (see Appendix 
\ref{App::AccuPSP}).

As becomes apparent from Eqs.\ \eqref{Eq::Cheb1} and \eqref{Eq::Cheb2}, the 
time evolution of the pure state $\ket{\psi(t)}$ requires the evaluation of 
matrix-vector products. Since ${\widetilde{\cal H}}$ is a sparse matrix, these 
matrix-vector multiplications can be implemented comparatively time and 
memory efficient. In particular, we here calculate the matrix elements of 
${\widetilde{\cal H}}$ \textit{on the fly} and use parallelization to 
reduce the runtime.
Thus, the memory requirements are 
essentially given by the size of the state $\ket{\psi(t)}$ and the auxiliary 
states $\ket{v_{k-1}}$, $\ket{v_{k}}$, and $\ket{v_{k+1}}$.
As a consequence, it is possible to treat system (or cluster) sizes 
significantly larger compared to full ED (here up to $28$ 
lattice sites 
with a Hilbert-space dimension of $d \approx 10^8$).   
Since the transverse-field Ising model \eqref{Eq::Hamiltonian} does 
not conserve the total 
magnetizations $X$ or $Z$, the corresponding quantum numbers cannot be used to 
block-diagonalize ${\cal 
H}$. Moreover, the clusters entering the NLCE are defined with open boundary 
conditions such that translational invariance cannot be exploited. 
Let us note that the clusters do have a reflection 
(parity) symmetry, which in principle can be used to reduce the memory 
requirements (though the reduction is less strong compared to the other 
symmetries mentioned before). In this paper, however, we do not exploit the 
reflection symmetry and always work in the full Hilbert space with dimension $d 
= 2^L$.

\section{Results}\label{Sec::Results}

We now present our numerical results for the quench dynamics of $\langle 
X(t) \rangle$ and $\langle 
Z(t) \rangle$ in chains, ladders, and two-dimensional lattices.
Our main focus is to analyze the convergence properties of the NLCE by 
comparing to direct simulations of finite systems with periodic boundary 
conditions (for open boundary conditions, see Appendix 
\ref{App::OBC}) and to 
existing data from the literature. 

\subsection{Chains}\label{Sec::Result_Chain}

The transverse-field Ising chain is a paradigmatic example of an exactly 
solvable model and analytical solutions have been known for a long time 
\cite{Pfeuty_1970, Barouch_1970, Barouch_1971, Barouch_1971_2} (see also 
Appendix \ref{App::1d}). 
Since quantum quenches in the Ising chain have been studied extensively before 
(see, 
e.g., Refs.\ \cite{Igl_i_2000, Sengupta_2004, Rossini_2009, Igl_i_2011, 
Foini_2011, Calabrese_2011, Calabrese_2012, Calabrese_2012_2}), the present 
section should be 
mainly understood as a consistency check for our numerical methods and a 
preparation for the study of ladders and two-dimensional lattices in Secs.\ 
\ref{Sec::Results_2LL} and \ref{Sec::Results_2D}. (It might be fair to say, 
however, that 
explicit visualizations of the analytical solutions, e.g., for the  
full time-dependent relaxation process of $\langle X(t) \rangle$ for 
specific initial states and transverse fields $g$, are
less often available in the literature.)     
\begin{figure}[tb]
 \centering 
 \includegraphics[width=\textwidth]{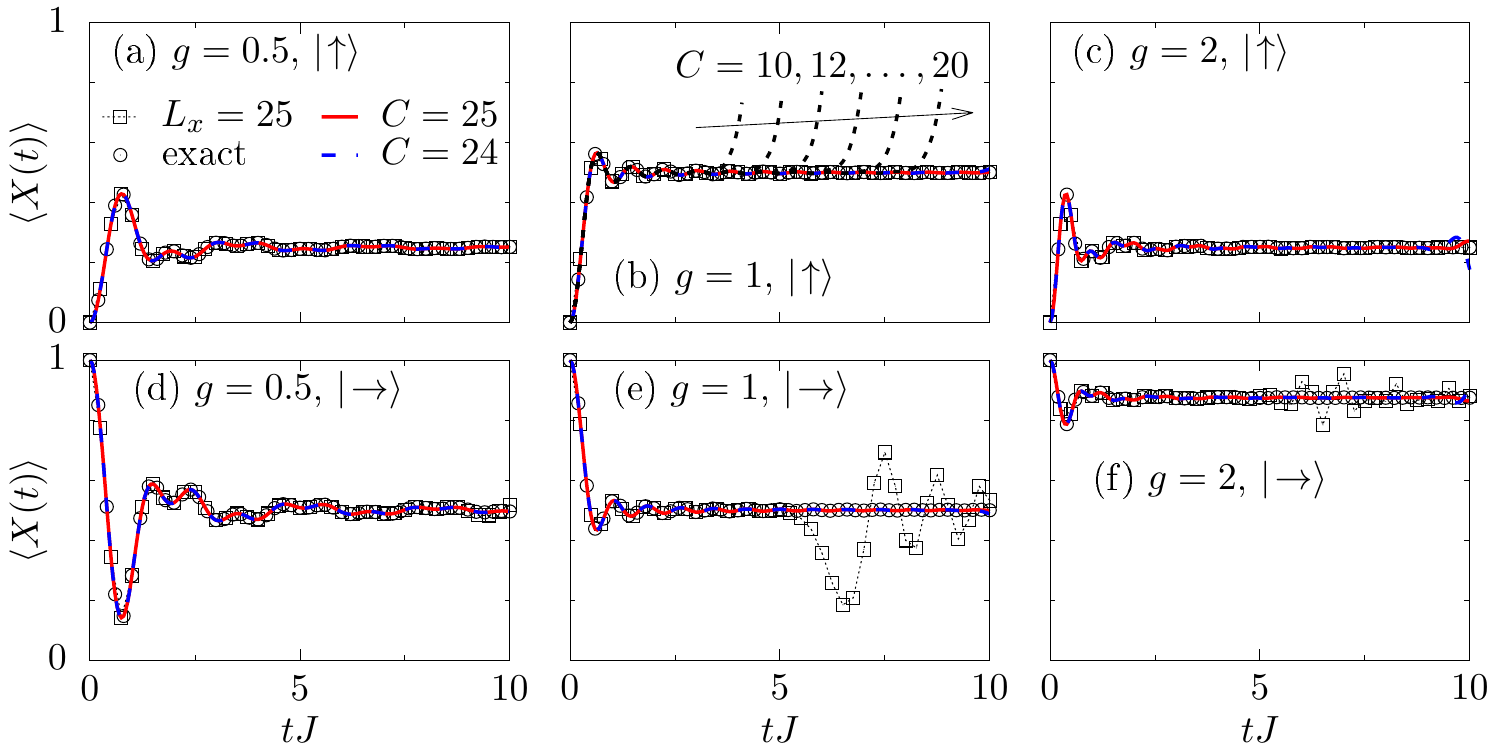}
 \caption{Dynamics of the transverse magnetization $\langle X(t) \rangle$ 
resulting from the initial state $\ket{\psi(0)} = \ket{\uparrow}$ [(a) - (c)], 
or $\ket{\psi(0)} = \ket{\rightarrow}$ [(d) - (f)], for chains with transverse 
fields $g = 0.5, 1, 2$. Numerical 
data obtained by NLCE for expansion orders $C = 24,25$ (blue 
and red curves) are compared to 
direct simulations for chains with $L_x = 25$ and PBC (open 
boxes), as well as to the 
exact, analytically known result  \cite{Barouch_1970, Barouch_1971, 
Barouch_1971_2} given in Eq.\ \eqref{Eq::ExactX} (open 
circles). In panel (b), we exemplarily show additional NLCE 
data for 
lower expansion orders $C = 10,12,\dots,20$.}
 \label{Fig2}
\end{figure}

In Figs.\ \ref{Fig2}~(a)-(c), the dynamics of the transverse 
magnetization 
$\langle X(t) \rangle$ is shown for quenches starting from the initial state 
$\ket{\psi(0)} = \ket{\uparrow}$ and different values of the transverse field 
$g = 0.5,1,2$. (Recall that the quantum critical point is $g = 1$ for the 
chain geometry.) Numerical data obtained by NLCE for expansion 
orders $C = 24, 25$ are compared to (i) a simulation for a finite 
chain with $L_x = 25$ and PBC, and (ii) the exact, analytically known result 
[see Eq.\ 
\eqref{Eq::ExactX} in Appendix \ref{App::1d}].
Note that we here choose to compare to systems with PBC, since 
finite-size effects are typically weaker in this case. For an additional 
comparison of NLCE results (also with lower expansion orders) to direct 
simulations of systems with OBC, see Appendix \ref{App::OBC}. 

Starting from its initial 
value 
$\langle X(0)\rangle = 0$, we find that the transverse 
magnetization 
$\langle X(t) \rangle$ in Figs.\ \ref{Fig2}~(a)-(c) quickly 
increases and exhibits a peak at short times, before equilibrating towards a 
constant long-time value. This stationary value is reached 
already for times $tJ \approx 2$. While this overall behavior of 
$\langle X(t) \rangle$ is very similar for all values of $g$ considered, the 
long-time value $\langle X(t \to \infty) \rangle$ is found to vary 
with $g$. In particular, it is known that this long-time value can be described 
in terms of a suitable GGE \cite{Essler_2016}. 

Generally, we find that the NLCE results in 
Figs.\ 
\ref{Fig2}~(a)-(c) are well converged on the time scales depicted, 
i.e., the curves for expansion orders $C = 24$ and $C = 25$ agree convincingly 
with each 
other. 
To visualize the convergence properties of the NLCE further, 
Fig.\ \ref{Fig2}~(b) shows additional NLCE data for lower expansion orders $C 
= 10,12,\dots,20$. Apparently, the convergence time of the expansion gradually 
increases with increasing $C$.
Furthermore, we find that the curves for the finite 
chain with 
$L_x = 25$ also nicely coincide with the NLCE 
data for $L \to \infty$, i.e., finite-site effects appear to 
be less relevant 
in these cases. Importantly, our numerical results for 
$\langle X(t) 
\rangle$ agree perfectly with the analytical solution. 

Next, in Figs.\ \ref{Fig2}~(d)-(f), we consider quenches starting from the 
state 
$\ket{\psi(0)} = \ket{\rightarrow}$. Despite the obvious difference that 
$\langle X(t)\rangle$ now starts at a maximum, $\langle X(0)\rangle = 
1$, the general picture is very 
similar compared to the 
previous case of $\ket{\psi(0)} = \ket{\uparrow}$. Namely, $\langle X(t) 
\rangle$ exhibits a
rapid decay and equilibrates rather quickly towards its long-time value.
Especially 
for $g = 1$ [Fig.\ \ref{Fig2}~(e)], however, we now observe pronounced  
finite-size effects, i.e., the curve for $L_x = 25$ deviates from the 
analytical solution for times $tJ \gtrsim 5$ and exhibits oscillations.
In contrast, the NLCE results for $C = 24, 25$ remain  
converged up to at least $tJ = 10$. This is a remarkable result since the 
largest cluster in 
the NLCE also only has $25$ lattice sites, i.e., the computational complexities 
of the NLCE and the simulation of the finite system are essentially the same.

Depending on the details of the quench, we thus find that performing a NLCE 
can yield a numerical advantage over the direct simulation of finite systems, 
see also 
Appendix \ref{App::OBC}.
On the one hand, if finite-size effects are weak, the results for finite 
chains can be very similar to the actual $L \to \infty$ dynamics (and also 
remain meaningful on longer 
time scales where the NLCE 
breaks down). On the other hand, the presence 
of strong finite-size 
effects [e.g.\ at the quantum critical point, cf.\ Fig.\ \ref{Fig2}~(e)] 
appears to favor the usage of NLCEs which yield the dynamics 
directly in the thermodynamic limit. This is a first result of the present 
paper. As will be discussed in more detail 
in the upcoming sections, a similar parameter-dependent advantage (or 
disadvantage) of performing a NLCE occurs for ladder geometries and 
two-dimensional lattices as well. 

\subsection{Ladders}\label{Sec::Results_2LL}

Let us now turn to the results for two- 
and three-leg ladders, which can be seen as intermediate cases between 
the chain geometry (cf.\ Sec.\ \ref{Sec::Result_Chain}) and the two-dimensional 
square lattice (cf.\ Sec.\ \ref{Sec::Results_2D}). Since exact solutions for 
the dynamics of ladders are
absent, we cannot compare our 
numerical data to analytical results. (For additional 
remarks on the transition from integrability to nonintegrability, see 
also Appendix \ref{App::EEE}.)    
\begin{figure}[tb]
 \centering 
 \includegraphics[width=1\textwidth]{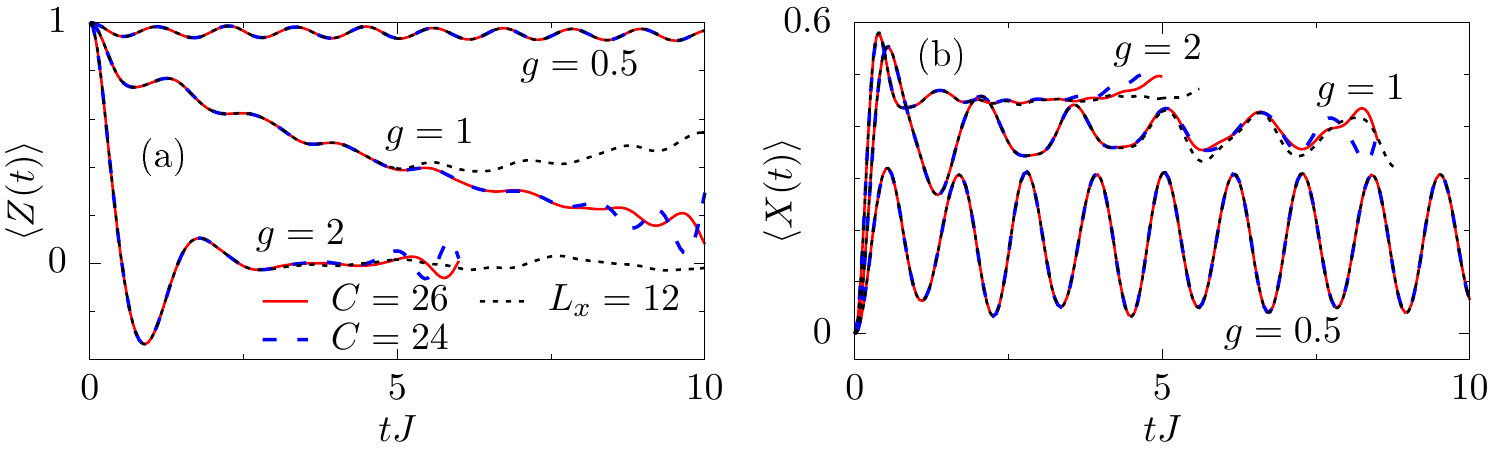}
 \caption{Dynamics of the (a) longitudinal magnetization 
$\langle Z(t) \rangle$ and (b) transverse magnetization 
$\langle X(t) \rangle$, in two-leg ladders with initial state 
$\ket{\psi(0)} 
= \ket{\uparrow}$ and different transverse fields 
$g$. Numerical 
data obtained by NLCE for expansion orders $C = 24,26$ are
compared to direct simulations for ladders with $L_x = 12$ and PBC.}
 \label{Fig3}
\end{figure}

In Fig.\ \ref{Fig3}, we consider quenches starting from the state 
$\ket{\psi(0)} = \ket{\uparrow}$ in two-leg ladders with different transverse 
fields $g$. Here, the data is obtained by NLCE for expansion orders $C = 24$ 
and $C = 26$, i.e., the largest clusters involved are of size $12 \times 2$ 
or $13 \times 2$.
As shown in Fig.\ \ref{Fig3}~(a), the dynamics of 
the longitudinal magnetization $\langle Z(t) \rangle$ displays a strong 
dependence on the value of $g$. On the one hand, for $g = 2$, $\langle Z(t) 
\rangle$ rapidly decays, exhibits a minimum at $tJ \approx 1$, and equilibrates 
to zero for $tJ \gtrsim 3$. On the other hand, for $g = 1$, the 
decay of $\langle Z(t) \rangle$ towards zero is distinctly slower 
and much more monotonous. Moreover, for $g = 0.5$ (i.e.\ a quench within the 
same 
equilibrium phase), the decay of $\langle Z(t) \rangle$ is almost 
indiscernible on the time scale shown, and we additionally observe that 
$\langle 
Z(t)\rangle$ exhibits small oscillations for this value of $g$.
The corresponding dynamics of the transverse magnetization $\langle X(t) 
\rangle$ is shown in Fig.\ \ref{Fig3}~(b).  
While $\langle X(t) \rangle$ quickly equilibrates towards a 
stationary value for $g = 2$, $\langle X(t) \rangle$ displays 
oscillations for $g = 0.5, 1$ which, especially in the case of $g = 
0.5$, do not equilibrate on the time scale 
shown here.

Let us comment on the convergence properties of the NLCE data in Fig.\ 
\ref{Fig3}. Both for $\langle Z(t)\rangle$ and $\langle X(t)\rangle$, we 
observe that the NLCE remains converged for longer times if the value of $g$ is 
smaller. Specifically, we find that the series breaks down at $tJ \approx 4$ 
for $g = 2$, at $tJ \approx 8$ for $g = 1$, while no breakdown can be seen for 
$g = 0.5$.  
Comparing these NLCE data to direct simulations of ladders with periodic 
boundary conditions and $L_x = 12$, a good agreement is found on short to 
intermediate time scales (or even longer for $g = 0.5$). 
In particular, the simulation for the finite ladder turns out to be 
advantageous for a strong transverse field $g = 2$, since it captures the 
stationary value of $\langle Z(t)\rangle$ and $\langle X(t)\rangle$ for a longer 
time than the NLCE. Similar to our previous results for chains, however, it 
becomes clear from Fig.\ \ref{Fig3}~(a) that the usage of NLCEs is 
in turn beneficial for $g = 1$, where finite-size effect appear to be stronger 
and the NLCE captures the monotonous decay of $\langle Z(t)\rangle$ up to
longer times compared to the finite-system data.

\begin{figure}[tb]
 \centering
 \includegraphics[width=1\textwidth]{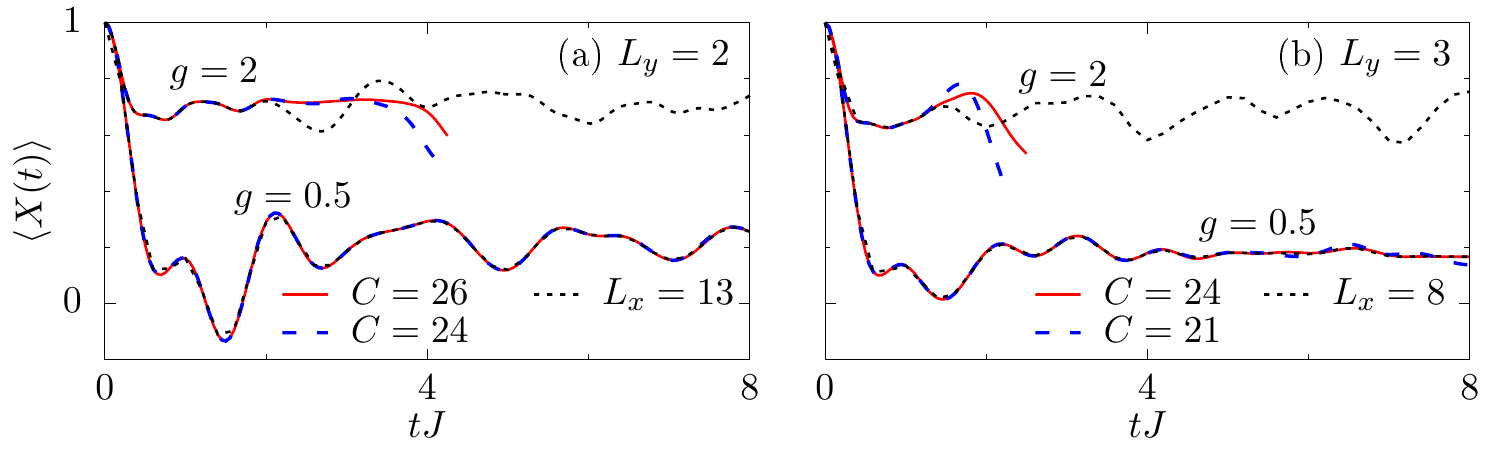}
 \caption{Dynamics of the transverse magnetization $\langle X(t) \rangle$ 
resulting from the initial state $\ket{\psi(0)} = \ket{\rightarrow}$ for (a) 
two-leg ladders and (b) three-leg ladders with $g = 0.5$ and $g = 2$. 
Numerical 
data obtained by NLCE for different expansion orders $C$ are compared to 
direct simulations of finite ladders with PBC. For 
additional NLCE data with lower expansion orders and a comparison to direct 
simulations of systems with OBC, see Appendix \ref{App::OBC}.}
 \label{Fig4}
\end{figure}

To proceed, Fig.\ \ref{Fig4} shows results for quantum quenches 
starting from the initial state $\ket{\psi(0)} = \ket{\rightarrow}$, with data 
for two-leg 
ladders in Fig.\ \ref{Fig4}~(a) and data for three-leg ladders in Fig.\ 
\ref{Fig4}~(b). Since $\langle Z(t) \rangle = 0$ due to the spin-flip symmetry 
of ${\cal H}$, 
we only have to consider $\langle X(t)\rangle$ in this case. We find that 
$\langle X(t) \rangle$ generally behaves very similar for the two different 
ladder geometries. Specifically, $\langle X(t) \rangle$ rapidly decays towards 
an (approximately constant) stationary value which is naturally higher for a 
higher value of $g$. Note however, that for $L_y = 2$ and $g = 0.5$, as well as 
for $L_y = 3$ and $g = 2$, $\langle X(t) \rangle$ still exhibits some residual 
fluctuations, i.e., perfect equilibration is absent.  
Concerning the convergence properties of the NLCE, we find that analogous to 
the previous case of $\ket{\psi(0)} = \ket{\uparrow}$ (cf.\ Fig.\ \ref{Fig3}), 
the NLCE remains 
converged significantly longer for $g = 0.5$ compared to $g = 2$. Especially the 
early breakdown of convergence for $L_y = 3$ and $g = 2$ in Fig.\ 
\ref{Fig4}~(b) emphasizes the fact that NLCEs are not necessarily the method of 
choice if one aims to study thermalization which typically requires 
the analysis of long time scales. (There  
exist, however, also examples where 
NLCEs yield converged results even in the infinite-time limit, 
i.e., converged results for the so-called diagonal ensemble
\cite{Rigol_2014}.) 
Eventually, let us note that the NLCE results and the data for 
systems with PBC in Fig.\ \ref{Fig4} yield a considerably longer convergence 
compared to analogous simulations 
for systems with OBC, see Appendix \ref{App::OBC} for details.

As a side remark to conclude the study of ladder geometries, let us note 
that 
Ref.\ \cite{vanVoorden_2020} has recently 
discussed the possibility of quantum scars in transverse-field Ising ladders. 
Specifically, Ref.\ \cite{vanVoorden_2020} has considered small 
values of $g$ and ``density-wave'' initial 
states of the form $\ket{\psi(0)} \sim \ket{\uparrow \downarrow \uparrow 
\downarrow \cdots}$.
These initial states were 
found to exhibit a large overlap with rare, weakly entangled eigenstates, 
leading to quasi-periodic revivals in the dynamics.
As detailed in Appendix \ref{App::EEE}, the fully polarized 
states $\ket{\uparrow}$ and $\ket{\rightarrow}$ studied in the present paper, 
in contrast, do not exhibit such a significant overlap with the weakly 
entangled
eigenstates. These special eigenstates therefore do not play a distinguished 
role for the quench dynamics presented in Figs.\ \ref{Fig3} and \ref{Fig4}. 

\subsection{Two-dimensional square lattice}\label{Sec::Results_2D}

We now come to the last part of this paper, i.e., the quantum quench dynamics 
in the two-dimensional transverse-field Ising model. Note that dynamical 
properties of this 
model 
\cite{Czarnik_2019,Hashizume_2018,Schmitt_2019,Guardado_Sanchez_2018, 
Hafner_2016,De_Nicola_2019}, as well as the emergence of thermalization 
\cite{Mondaini_2016, 
Mondaini_2017, Bla__2016}, have been studied before by a variety of approaches. 
By comparing our results to existing data from the 
literature, let us demonstrate in this section that numerical linked 
cluster expansions based on rectangular clusters only, combined with an 
efficient forward 
propagation of pure states, provide a competitive alternative to other 
state-of-the-art numerical approaches.  
\begin{figure}[tb]
 \centering 
 \includegraphics[width=1\textwidth]{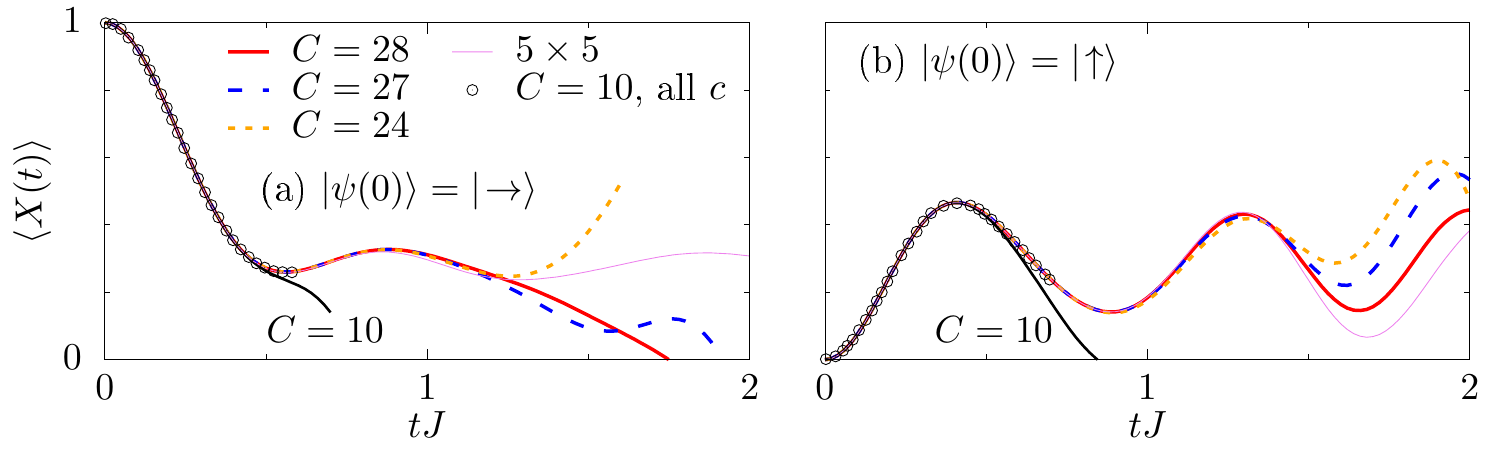}
 \caption{Dynamics of the transverse magnetization $\langle X(t) \rangle$ for a 
two-dimensional square lattice with transverse field $g = 1$,
obtained by NLCE with rectangular clusters and expansion orders $C = 24, 27, 
28$. 
The open circles are NLCE data digitized from Ref.\ 
\cite{White_2017}, where all (also nonrectangular) cluster geometries with 
up to $10$ lattice sites have been  
considered. In addition, we present data from a rectangle 
expansion up to $C = 10$ which, in comparison, converges to slightly shorter 
times than the full expansion.
The dynamics for a $5\times 5$ lattice with PBC is shown as 
well. The initial state is chosen as (a) $\ket{\psi(0)} = \ket{\rightarrow}$ 
and (b) 
$\ket{\psi(0)} = 
\ket{\uparrow}$.}
 \label{Fig5}
\end{figure}

As a first step, it is instructive to compare our results to 
earlier NLCE data from Ref.\ \cite{White_2017}. 
This comparison is shown in Figs.\ 
\ref{Fig5}~(a) and (b), where the dynamics of the transverse 
magnetization $\langle X(t) \rangle$ is studied for quenches from 
$\ket{\rightarrow}$ and $\ket{\uparrow}$ with $g = 1$. (Recall that $g_c 
\approx 3.044$ for the two-dimensional lattice.) Importantly, Ref.\ 
\cite{White_2017} 
has considered all (also nonrectangular) cluster 
geometries 
in the expansion and has used full ED to evaluate the 
respective weights. Due to the 
computational bottlenecks of NLCEs discussed in Sec.\ \ref{Sec::Numerics_NLCE}, 
Ref.\ \cite{White_2017} was consequently limited to rather small clusters with 
up to 
$10$ lattice sites. In Fig.\ \ref{Fig5}, we find that our
NLCE with solely rectangular clusters nicely reproduces the data from Ref.\ 
\cite{White_2017}. 
In particular, while the results of  
Ref.\ \cite{White_2017} are converged for times $tJ < 
1$, the rectangular NLCE up to expansion order $C = 28$ 
(i.e.\ the 
largest 
clusters are of size $7 \times 4$, $14 \times 2$, $28\times 1$) yields 
converged results on time scales which are approximately twice as long.
This demonstration, that a NLCE restricted to rectangular cluster geometries 
can be better than a NLCE comprising all (possibly nonrectangular) clusters,  
is an important 
result of the present paper. 

Let us add some comments on the convergence properties of the 
NLCE in Fig.\ \ref{Fig5}. First, as an additional comparison between the
rectangle expansion and the full expansion from Ref.\ \cite{White_2017}, Figs.\ 
\ref{Fig5}~(a) and (b) also show data obtained by the rectangle expansion 
with the lower expansion order $C = 10$. For this value of $C$, we find that 
the rectangle expansion is converged to slightly shorter times than the 
data from Ref.\ \cite{White_2017}. This is expected since, for a fixed value of 
$C$, the full expansion should always perform equally well or better compared to 
any restricted NLCE. However, let us stress once again the crucial advantage of
the rectangle expansion that higher 
expansion orders can be included due to the reduced 
combinatorial costs. 
Secondly, we note that given the NLCE results 
up to expansion order  $C = 28$ in Figs.\ \ref{Fig5}~(a) and (b), the 
short-time dynamics for this value of the transverse field can 
apparently be accessed also by 
the direct simulation of a $5 \times 5$ lattice with PBC. This 
is similar to our previous findings for chains and ladders in Secs.\ 
\ref{Sec::Result_Chain} and \ref{Sec::Results_2LL}. Namely, depending on the 
parameter regime, NLCEs might not necessarily outperform a direct simulation of 
a finite system with PBC if the latter yields small finite-size effects. As 
shown in Appendix \ref{App::OBC}, however, the advantage of the NLCE is more 
pronounced when one compares to direct simulations of systems with OBC 
instead.
\begin{figure}[tb]
 \centering 
 \includegraphics[width=1\textwidth]{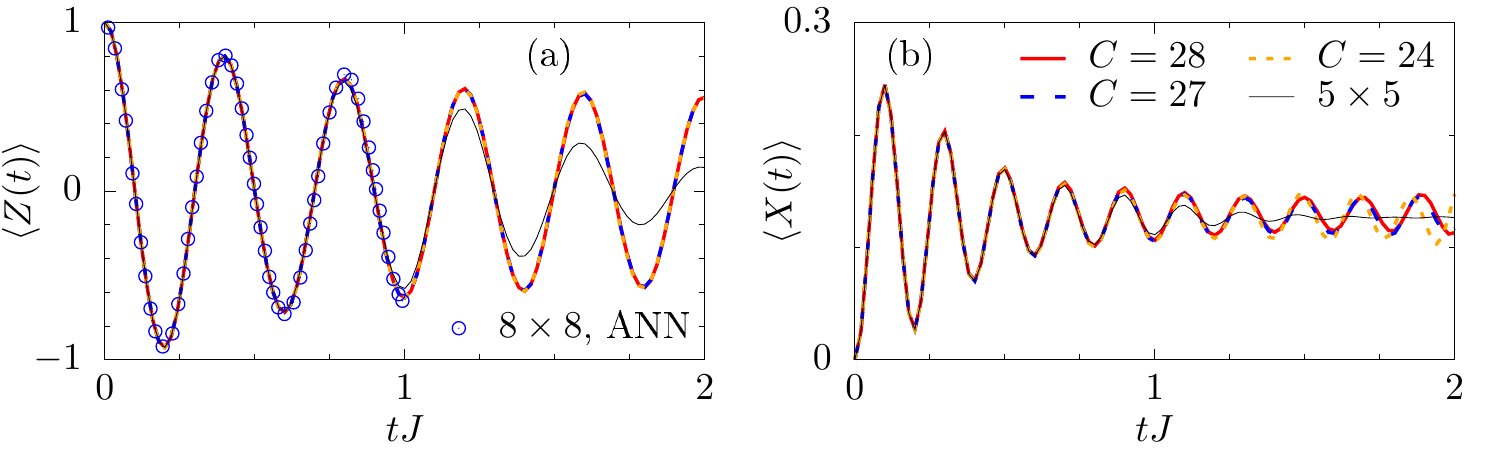}
 \caption{Dynamics of the (a) longitudinal magnetization 
$\langle Z(t) \rangle$ and (b) transverse magnetization 
$\langle X(t) \rangle$, resulting from the initial state 
$\ket{\psi(0)} 
= \ket{\uparrow}$ for a two-dimensional square lattice with $g = 
2.63g_c$. Data obtained by NLCE for expansion orders $C = 24, 27, 28$ are
compared to a simulation of a $5 \times 5$ lattice with PBC. 
In (a), we additionally show digitized ANN data from Ref.\ \cite{Schmitt_2019} 
for a $8 \times 8$ lattice.}
 \label{Fig6}
\end{figure}

Next, let us study quenches starting from the state $\ket{\psi(0)} = 
\ket{\uparrow}$ such that $\langle Z(0) \rangle = 1$ and $\langle X(0)\rangle = 
0$, and consider a strong transverse field $g = 2.63g_c \approx 8$, i.e., a 
quench across 
the quantum critical point.  Again, we consider clusters 
with up to $28$ lattice sites in the NLCE. 
In Fig.\ \ref{Fig6}~(a), we find that $\langle Z(t) \rangle$ displays 
pronounced oscillations with an amplitude that is weakly damped over time. 
Correspondingly, the transverse magnetization $\langle X(t) \rangle$ in 
Fig.\ \ref{Fig6}~(b) exhibits damped oscillations as well (with a frequency 
that is twice as 
large). It is instructive to compare these NLCE data for the thermodynamic 
limit to a simulation of a $5\times 5$ lattice with PBC. 
Specifically, 
one observes that for such a finite 
system and times $tJ \gtrsim 1$, the oscillations of $\langle Z(t) 
\rangle$ and $\langle X(t)\rangle$ die away rather quickly. This is in contrast 
to the NLCE results for $L \to \infty$ which capture the persistent 
oscillations on a longer time scale.    
In addition, we compare our NLCE results for $\langle Z(t)\rangle$ in Fig.\ 
\ref{Fig6}~(a) to recent data digitized from Ref.\ \cite{Schmitt_2019}, which 
are computed by an artificial neural-network (ANN) approach for a $8\times 8$ 
lattice. While the NLCE and ANN data agree nicely with each other 
for 
times $tJ < 1$, the NLCE remains converged also on longer time 
scales. In particular,  
the ANN data from Ref.\ \cite{Schmitt_2019} up to times 
$tJ \lesssim 1$ can be reproduced even by the 
smaller $5\times 5$ lattice. Thus, for the parameter regime considered in Fig.\ 
\ref{Fig6}, it appears that the NLCE
can be better than the direct simulation of finite systems with PBC as well as 
the ANN 
approach from Ref.\ \cite{Schmitt_2019}.  
This is another important result of the present paper.

\begin{figure}[tb]
 \centering 
 \includegraphics[width=\textwidth]{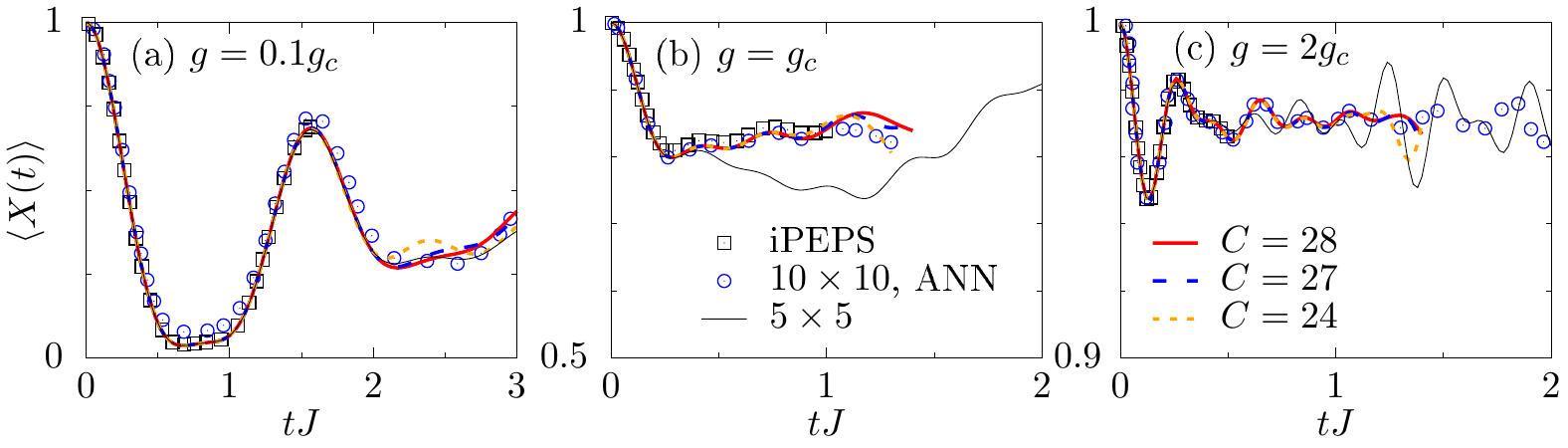}
 \caption{Dynamics of the transverse magnetization $\langle X(t) \rangle$ for 
two-dimensional lattices with initial state $\ket{\psi(0)} = 
\ket{\rightarrow}$ and transverse fields (a) $g = 0.1g_c$, (b) $g = 1g_c$, 
and (c) $g = 2g_c$. Data obtained by NLCE for expansion orders $C = 
24,27,28$ are compared to the simulation of a $5 \times 5$ lattice with 
PBC. Additionally, we show iPEPS data digitized from Ref.\ \cite{Czarnik_2019} 
and ANN data for a $10\times 10$ lattice digitized from Ref.\ 
\cite{Schmitt_2019}.}
 \label{Fig7}
\end{figure}

Finally, we also consider quenches starting from the state $\ket{\psi(0)} = 
\ket{\rightarrow}$. The values of the transverse 
field are chosen as $g = 0.1 g_c, 1g_c, 2g_c$, which again allows us to 
compare to ANN data from Ref.\ \cite{Schmitt_2019}, as well as to 
data from 
Ref.\ \cite{Czarnik_2019} based on infinite projected entangled pair states 
(iPEPS). For all values of $g$ shown in Figs.\ \ref{Fig7}~(a)-(c), we find a 
convincing agreement between the data from Refs.\ \cite{Czarnik_2019, 
Schmitt_2019} and our NLCE results up to expansion order $C = 28$, with 
convergence times 
that are rather similar 
for all three methods.
In order to put the convergence times into 
perspective, 
it is again helpful to compare the NLCE data  to a simulation of a finite 
$5\times 5$ lattice with PBC. While 
finite-size effects appear to be less important for $g = 0.1g_c$ and $g 
= 
2g_c$, we observe 
pronounced finite-size effects for $g = g_c$ already 
at short times $tJ \approx 0.5$ due to, e.g., the 
divergence of the relevant length scales at the quantum critical point. 
Importantly, the NLCE results for $g = g_c$ in Fig.\ \ref{Fig7}~(b) remain 
converged up to times $tJ \approx 1.5$.
One explanation for the advantage of NLCEs at the quantum critical point might 
be given by the fact that the expansion involves a variety of clusters with 
different ratios of width and height such 
that one can capture the dynamics on longer time and length  
scales. This is another central result of this paper. In this context, let us 
add that the inclusion of rectangles with different 
length ratios appears to be crucial to achieve a good convergence. For 
instance, we have checked that an expansion using solely square-shaped 
clusters ($1\times 1, 2\times2,\dots,5\times5$) 
performs very poorly instead (not shown here).

\section{Conclusion}\label{Sec::Conclusion}

To summarize, we have studied the nonequilibrium dynamics of the transverse and 
the longitudinal magnetization resulting from quantum quenches with fully 
polarized 
initial states in the transverse-field Ising model defined on different lattice
geometries. To this end, we have relied on an efficient combination of 
numerical linked 
cluster expansions and a forward propagation of 
pure states via 
Chebyshev polynomials. 

Depending on the geometry and the parameter regime under consideration, the 
quench dynamics has been found to display a variety of different 
behaviors ranging 
from quick equilibration, over slower 
monotonous relaxation, to persistent (weakly damped) oscillations. 
As a main result, we have demonstrated that NLCEs comprising solely rectangular 
clusters provide a promising approach 
to study the dynamics of two-dimensional quantum many-body systems directly in 
the thermodynamic limit. While the 
organization of the NLCE becomes straightforward due to the simple cluster 
geometry, the memory efficient pure-state propagation made it possible to 
include clusters with up to 
$28$ lattice sites. Especially, for quenches to the quantum critical point, 
where finite-size effects are typically strong, we have shown that NLCEs can 
yield converged results on time scales 
which compare favorably to direct simulations of finite systems with periodic 
boundary conditions (also in the 
case of chains or ladders). By comparing to existing data from 
the literature, we have demonstrated that the reachable time scales are 
also competitive to other state-of-the-art 
numerical methods. While NLCEs with rectangular clusters have been used before 
to 
obtain thermodynamic quantities \cite{Bruognolo_2017} or entanglement entropies 
\cite{Kallin_2013}, the present paper unveils that such NLCEs also provide a 
powerful 
tool to study the real-time dynamics of quantum 
many-body systems.  

A natural direction of future research is to further explore 
the capabilities of NLCEs to simulate quantum quench dynamics of 
two-dimensional systems in the thermodynamic limit.  
In this context, it might be promising to consider other building blocks for 
the expansion such as, e.g., clusters that consist of multiple corner-sharing 
$2\times 
2$ squares \cite{Bhattaram_2019}.
Moreover, it will be interesting to study
other two-dimensional 
lattice geometries 
such as triangular or Kagome lattices with nonrectangular cluster shapes.  
One particular question in the field of quantum many-body 
dynamics where NLCEs might be able to contribute 
is the existence of many-body localization in higher dimensions. While 
the usage of NLCEs in disordered systems involves additional complications 
beyond our explanations in Sec.\ \ref{Sec::Numerics_NLCE}, NLCEs can yield 
results directly in the thermodynamic limit which is especially important close 
to the potential transition between the thermal and the MBL regime.
Although truly long times might 
still remain out of reach, the usage of supercomputing will be helpful to 
include higher 
expansion orders (up to $C \approx 
40$ \cite{Richter_2019_2}), which improves the 
convergence of the NLCE even
further.

\textit{Note added:} After this paper was submitted, we became aware of the 
related work \cite{Gan_2020} which appeared in the same arXiv posting as our 
manuscript. While Ref.\ \cite{Gan_2020} also presents NLCE calculations for 
the dynamics of two-dimensional systems using an expansion in rectangles, its 
focus is on the application of NLCEs to disordered systems and inhomogeneous 
initial states. In addition, while Ref.\ \cite{Gan_2020} employs full
ED to evaluate the contributions of the clusters, the present paper highlights 
the usefulness of efficient pure-state propagation methods to reach 
expansion orders beyond the range of full ED and to extend the 
convergence times of 
the NLCE. 

\section*{Acknowledgements}

The authors thank F.\ Jin for very helpful discussions.

\paragraph{Funding information}

This work has been funded by the Deutsche
Forschungsgemeinschaft (DFG) - Grants No. 397067869
(STE 2243/3-1), No. 355031190 - within the DFG Research Unit FOR 2692.

\appendix

\section{Exact solution for the integrable chain}\label{App::1d}

In the case of a chain geometry, the transverse-field Ising model 
\eqref{Eq::Hamiltonian} is a paradigmatic example of an integrable model and 
can 
be 
diagonalized by means of subsequent Jordan-Wigner, Fourier, and Bogolioubov 
transforms \cite{Pfeuty_1970}, 
\begin{equation}\label{Eq::HDiag}
 {\cal H} = \sum_{k} E_k \eta_k^\dagger \eta_k + \text{const.}\ ,\quad E_k 
= 2J\sqrt{(g-\cos k)^2+ \sin^2k}\ .
\end{equation}
Since quantum quenches in the transverse-field Ising chain have been studied 
extensively before, and since the focus of this paper is 
on the numerical analysis 
of nonintegrable geometries, we here refrain from 
providing more details and refer to the large body of existing 
literature instead \cite{Igl_i_2000, Sengupta_2004, Rossini_2009, Igl_i_2011, 
Foini_2011, Calabrese_2011, Calabrese_2012, Calabrese_2012_2}.  
Given the notation of ${\cal H}$ in Eqs.\ \eqref{Eq::Hamiltonian} and 
\eqref{Eq::HDiag}, as well as an initial state $\ket{\psi(0)}$ which is chosen 
as the groundstate of ${\cal H}$ for some transverse field $g^\prime$,  
the 
dynamics of the transverse magnetization $\langle 
X(t) \rangle$ for a quench $g^\prime \to g$ is then given by 
\cite{Barouch_1970, Barouch_1971, Barouch_1971_2, Sengupta_2004}, 
\begin{equation}\label{Eq::ExactX}
 \langle X(t) \rangle = 2\int_0^\pi \frac{\text{d}k}{2\pi} \frac{1}{E_k^2 
E_k^\prime} \left[\epsilon_k(\epsilon_k \epsilon_k^\prime + \gamma_k^2)+ 
\gamma_k^2 (\epsilon_k^\prime - \epsilon_k)\cos(2 E_k t)\right]\ , 
\end{equation}
where we have used the abbreviations 
\begin{equation}
 \epsilon_k = 2J(g-\cos k)\ 
,\quad  \gamma_k = 2J\sin k\ ,
\end{equation}
and $E_k^\prime$ and $\epsilon_k^\prime$ are defined like their unprimed 
counterparts, but with $g \to g^\prime$. 
In order to obtain 
the results shown in Fig.\ \ref{Fig2} of the main text, we 
have numerically evaluated the integral in Eq.\ \eqref{Eq::ExactX} either for 
$g^\prime = 0$ ($\ket{\psi(0)} = \ket{\uparrow}$) or for $g^\prime \to \infty$ 
($\ket{\psi(0)} = \ket{\rightarrow}$).

\section{Accuracy of the pure-state propagation}\label{App::AccuPSP}
\begin{figure}[tb]
 \centering
 \includegraphics[width=1\textwidth]{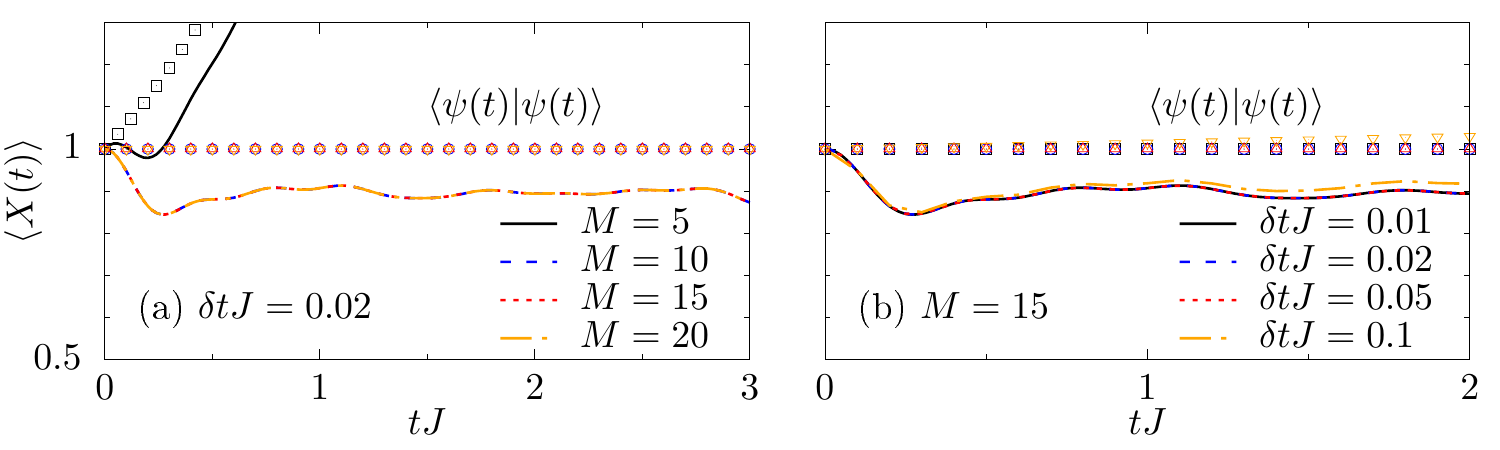}
 \caption{Dynamics of the transverse magnetization $\langle X(t) \rangle$ for a 
cluster of size $L_x \times L_y = 7 
\times 3$ (with OBC), initial state $\ket{\psi(0)} = \ket{\rightarrow}$,  and 
transverse field $g = 3.044$. The symbols indicate the norm $\langle 
\psi(t)|\psi(t)\rangle$. (a) Fixed time step $\delta t J = 0.02$ and varying 
expansion order $M$. (b) Fixed $M = 15$ and varying $\delta t J$.}
 \label{Fig8}
\end{figure}

While we have already demonstrated that our numerical results agree very well 
with existing data, let us nevertheless discuss the accuracy 
of the Chebyshev-polynomial expansion which is used to evaluate the time 
evolution of the pure states $\ket{\uparrow}$ and $\ket{\rightarrow}$. To this 
end, Fig.\ \ref{Fig8}, shows the dynamics of the transverse magnetization 
$\langle X(t) \rangle$ for a cluster of size $L_x \times L_y = 7 
\times 3$ (with OBC), initial state $\ket{\psi(0)} = \ket{\rightarrow}$, and 
transverse field $g = g_c \approx 3.044$.

First, in Fig.\ \ref{Fig8}~(a), we 
set the 
discrete time step to $\delta t J = 0.02$ and depict results for different 
expansion orders $M = 5,10,15,20$ (curves). On the one hand, for small $M = 5$, 
we 
observe clearly unphysical results (e.g.\ $\langle X(t)\rangle > 1$), which can 
also be explained by the fact that the norm $\langle \psi(t)|\psi(t)\rangle$ 
(symbols) is 
not conserved over time for this choice of $M$. On the other 
hand, for $M = 10,15,20$, all curves for $\langle X(t)\rangle$ are perfectly on 
top of each other, i.e., convergence with respect to $M$ has been reached, and 
$\langle \psi(t)|\psi(t)\rangle = 1$.  

Next, Fig.\ \ref{Fig8}~(b) shows results for a fixed expansion order $M = 
15$ and varying time step $\delta t J = 0.01, 0.02, 0.05, 0.1$. We find that 
$\langle X(t)\rangle$ is practically independent of the time step for the 
three smallest values of $\delta t J$ used here. However, visible 
deviations occur in the case of the largest time step $\delta t J = 
0.1$. 
While the required time step $\delta t$ and expansion order $M$ can certainly 
depend on the 
parameter regime under consideration, the typical choice used in the main 
text, i.e., $\delta t J = 0.02$ and $M = 15$, appears to yield very accurate 
results.

\section{Additional results for lower expansion orders and 
systems with open boundary conditions}\label{App::OBC}
\begin{figure}[tb]
 \centering
 \includegraphics[width=1\textwidth]{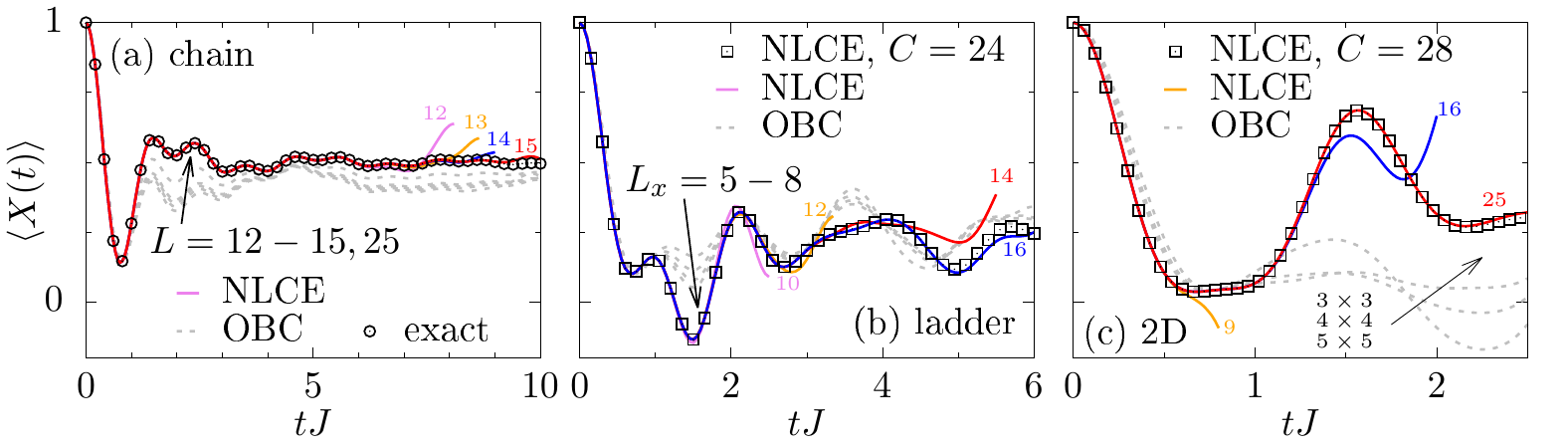}
 \caption{Dynamics of the tranverse magnetization $\langle X(t) 
\rangle$ for quenches in (a) chains with $g = 0.5$, (b) two-leg laddders ($L = 
L_x \times 2$) 
with $g = 0.5$, and (c) two-dimensional lattices with $g = 0.1 g_c$. The 
initial 
state is $\ket{\psi(0)} = \ket{\rightarrow}$. Results obtained by numerical 
linked cluster expansion for different expansion orders $C$ (solid 
curves) are compared to 
direct simulations for systems with open boundary conditions (gray dashed 
curves). In all cases, we find that for a given expansion order $C$ (or system 
size $L$) the NLCE yields converged results for significantly longer times than 
the corresponding direct simulation with OBC.   
}
 \label{Fig9}
\end{figure}

In Figs.\ \ref{Fig2} - \ref{Fig7} of the main text, we have 
compared the convergence of the NLCE to direct simulations of systems with 
periodic boundary conditions. However, since the clusters entering the NLCE are 
defined with open 
boundary conditions, it might be interesting to compare the convergence of the 
NLCE to direct simulations with OBC as well. Such a comparison is shown in 
Fig.\ \ref{Fig9} for the dynamics of the transverse magnetization $\langle X(t) 
\rangle$ in chains [panel (a)], ladders [panel (b)], and 
two-dimensional square lattices [panel (c)], with one exemplarily chosen 
transverse field $g$ in each case. Specifically, the curves 
shown in Fig.\ \ref{Fig9} are complementary to our earlier data in Figs.\ 
\ref{Fig2}~(d), \ref{Fig4}~(a), and \ref{Fig7}~(a), as we now also include NLCE 
results for lower expansion orders. 
Moreover, to guarantee a fair comparison, Fig.\ \ref{Fig9} always shows curves 
for matching system sizes and expansion orders, i.e., $L = C$ (recall that the 
expansion order of the NLCE is defined as the largest cluster size involved in 
the expansion). Importantly, we find that the NLCE yields converged results on 
significantly longer time scales than the simulation of the finite system with 
OBC for all cases shown here. For instance, in the case of the chain [Fig.\ 
\ref{Fig9}~(a)], expansion order $C = 15$ is already sufficient to yield 
converged results up to $tJ = 10$, whereas the direct simulation for a system 
with OBC fails to capture the correct long-time plateau even for the 
considerably larger system size $L = 25$. Similarly, in the case of the two-leg 
ladder [Fig.\ 
\ref{Fig9}~(b)], the curves for finite systems with OBC converge only up to 
the rather short time $tJ \approx 1$, while the convergence of the NLCE 
quickly improves with increasing $C$. 
Especially for the two-dimensional case [Fig.\ 
\ref{Fig9}~(c)], the simulations for the finite system with OBC even fail to 
describe the initial decay of $\langle X(t) \rangle$ correctly.   
Thus, we conclude that for a given expansion order $C$ (or system size $L$) the 
NLCE performs considerably better than the corresponding direct simulation 
with OBC.

\section{Eigenstate entanglement and spectral decomposition of initial 
states}\label{App::EEE}
\begin{figure}[tb]
\floatbox[{\capbeside\thisfloatsetup{capbesideposition={right,top},
capbesidewidth=0.3\textwidth}}]{figure}[\FBwidth]
{\caption{(a) and (b) Eigenstate entanglement entropy $S_{\ket{n}}$ of a chain 
and a two-leg ladder. The dashed line indicates the ``Page value'' for a 
random state \cite{Page_1993}, while the dotted line indicates the maximum 
entropy possible for the chosen bipartition. 
(c) and (d) Overlap of initial states 
$\ket{\uparrow}$ and  
$\ket{\rightarrow}$ with the eigenstates $\ket{n}$. We have $L = 14$, $g = 
0.5$, 
and PBC in all cases.}
\label{Fig10}}
{\includegraphics[width=0.7\textwidth]{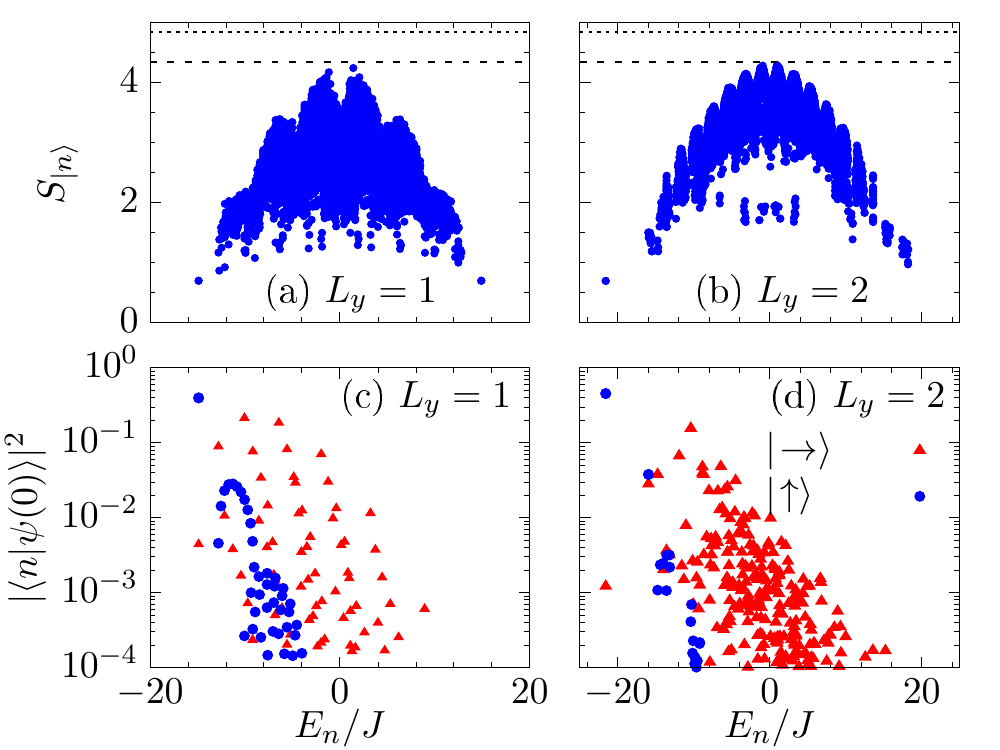}}
\end{figure}

Let us discuss some properties of the fully 
polarized initial states $\ket{\uparrow}$ and $\ket{\rightarrow}$. 
To this end, we first study 
the 
entanglement 
(von Neumann) entropy $S_{\ket{n}}$ of the eigenstates $\ket{n}$ of ${\cal 
H}$, 
\begin{equation}
 S_{\ket{n}} = -\text{Tr}[\rho_A \ln \rho_A]\ ,\quad \rho_A = 
\text{Tr}_B\left\lbrace\ket{n}\bra{n}\right\rbrace\ , 
\end{equation}
where $\rho_A$ is the reduced density matrix on a subsystem $A$, obtained 
by 
tracing over the degrees of freedom in the complement $B$. 
In Figs.\ \ref{Fig10}~(a) and (b), $S_{\ket{n}}$ is shown for a chain and a 
two-leg ladder respectively, numerically obtained by full 
exact diagonalization for 
$L = 14$ sites, transverse field $g = 0.5$, and periodic boundary conditions. 
In both cases, we have chosen $A$ as one half of the system, i.e., the first 
$7$ lattice sites in case of the chain, or the first three rungs and one site 
of the fourth rung in case of the ladder.  
On the one hand, for the integrable chain geometry in Fig.\ \ref{Fig10}~(a), we 
find that 
$S_{\ket{n}}$ is 
comparatively small at the edges (consistent with  
the area-law entanglement scaling of groundstates \cite{Eisert_2010}), 
while weakly and 
strongly entangled states coexist in the bulk of the spectrum (see also Refs.\ 
\cite{Alba_2009, Beugeling_2015}).
On the other hand, for the two-leg ladder in Fig.\ \ref{Fig10}~(b), 
the fluctuations of $S_{\ket{n}}$ in the center of the spectrum are clearly 
smaller, i.e., the eigenstates are
typically stronger entangled. This behavior of $S_{\ket{n}}$ can be 
interpreted as an indication of the transition from integrability to 
nonintegrability \cite{Beugeling_2015}, by going from chains to ladders. 
In addition, we can identify a small number of 
eigenstates $\ket{n}$ with energy close to $E = 0$ in Fig.\ \ref{Fig10}~(b), 
which exhibit a distinctly 
lower value of $S_{\ket{n}}$. This appears to be consistent with the recent 
proposal of quantum scars in transverse-field Ising ladders in 
Ref.\ \cite{vanVoorden_2020}.

Next, it is useful to study 
$S_{\ket{n}}$ in combination with the overlap between the initial states 
$\ket{\psi(0)} = \ket{\uparrow}, \ket{\rightarrow}$ and the eigenstates 
$\ket{n}$, 
\begin{equation}
P_{\ket{\psi}} = \sum_{n=1}^D |\langle n|\psi(0)\rangle|^2 \delta(E-E_n)\ , 
\end{equation}
where $E_n$ is the eigenvalue of ${\cal H}$ belonging to $\ket{n}$.
As shown in Figs.\ \ref{Fig10}~(c) and (d), this spectral distribution is 
narrow 
and peaked at the groundstate in the case of $\ket{\uparrow}$, 
while $ P_{\ket{\psi}}$ is much broader for $\ket{\rightarrow}$, both for the 
chain and the ladder. 
Thus, a quench to $g = 0.5$ with $\ket{\psi(0)} = \ket{\uparrow}$, results in 
a dynamics which is strongly dominated 
by the groundstate with a significantly smaller admixture of excited states. 
Note 
that exactly for such a situation, i.e., a quantum 
many-body system with one macroscopically populated eigenstate, an analytical 
prediction for the temporal relaxation has been recently obtained in Ref.\ 
\cite{Reimann_2020}. While this is beyond the scope of the present manuscript, 
it appears that 
quantum quenches in transverse-field Ising 
chains or ladders can be promising candidates to test such 
predictions.

Finally, as already pointed out in the main text, 
we note that the fully polarized initial states $\ket{\uparrow}$ and 
$\ket{\rightarrow}$ do not exhibit a distinguished overlap with the rare, 
weakly entangled eigenstates discussed in Fig.\ \ref{Fig10}~(b). These 
potential 
quantum scars therefore do not play an essential role for the resulting quench 
dynamics.

\bibliography{NLCE_2D}

\begin{thebibliography}{100}
\providecommand{\url}[1]{\texttt{#1}}
\providecommand{\urlprefix}{URL }
\expandafter\ifx\csname urlstyle\endcsname\relax
  \providecommand{\doi}[1]{doi:\discretionary{}{}{}#1}\else
  \providecommand{\doi}{doi:\discretionary{}{}{}\begingroup
  \urlstyle{rm}\Url}\fi
\providecommand{\eprint}[2][]{\url{#2}}

\bibitem{Polkovnikov_2011}
A.~Polkovnikov, K.~Sengupta, A.~Silva and M.~Vengalattore,
\newblock \emph{Colloquium: Nonequilibrium dynamics of closed interacting
  quantum systems},
\newblock Rev. Mod. Phys. \textbf{83}(3), 863 (2011),
\newblock \doi{10.1103/revmodphys.83.863}.

\bibitem{Eisert_2015}
J.~Eisert, M.~Friesdorf and C.~Gogolin,
\newblock \emph{Quantum many-body systems out of equilibrium},
\newblock Nat. Phys. \textbf{11}(2), 124 (2015),
\newblock \doi{10.1038/nphys3215}.

\bibitem{Bertini2020}
B.~Bertini, F.~Heidrich-Meisner, C.~Karrasch, T.~Prosen, R.~Steinigeweg and
  M.~{\v{Z}}nidari{\v{c}},
\newblock \emph{Finite-temperature transport in one-dimensional quantum lattice
  models} \eprint{https://arXiv.org/abs/2003.03334}.

\bibitem{Mitra_2018}
A.~Mitra,
\newblock \emph{Quantum quench dynamics},
\newblock Annu. Rev. Condens. Matter Phys. \textbf{9}(1), 245 (2018),
\newblock \doi{10.1146/annurev-conmatphys-031016-025451}.

\bibitem{Dziarmaga_2010}
J.~Dziarmaga,
\newblock \emph{Dynamics of a quantum phase transition and relaxation to a
  steady state},
\newblock Adv. Phys. \textbf{59}(6), 1063 (2010),
\newblock \doi{10.1080/00018732.2010.514702}.

\bibitem{Reimann_2016}
P.~Reimann,
\newblock \emph{Typical fast thermalization processes in closed many-body
  systems},
\newblock Nat. Commun. \textbf{7}(1), 10821 (2016),
\newblock \doi{10.1038/ncomms10821}.

\bibitem{Erne_2018}
S.~Erne, R.~Bücker, T.~Gasenzer, J.~Berges and J.~Schmiedmayer,
\newblock \emph{Universal dynamics in an isolated one-dimensional bose gas far
  from equilibrium},
\newblock Nature \textbf{563}(7730), 225 (2018),
\newblock \doi{10.1038/s41586-018-0667-0}.

\bibitem{Pr_fer_2018}
M.~Prüfer, P.~Kunkel, H.~Strobel, S.~Lannig, D.~Linnemann, C.-M. Schmied,
  J.~Berges, T.~Gasenzer and M.~K. Oberthaler,
\newblock \emph{Observation of universal dynamics in a spinor bose gas far from
  equilibrium},
\newblock Nature \textbf{563}(7730), 217 (2018),
\newblock \doi{10.1038/s41586-018-0659-0}.

\bibitem{Richter_2019_3}
J.~Richter and R.~Steinigeweg,
\newblock \emph{Relation between far-from-equilibrium dynamics and equilibrium
  correlation functions for binary operators},
\newblock Phys. Rev. E \textbf{99}(1), 012114 (2019),
\newblock \doi{10.1103/physreve.99.012114}.

\bibitem{Richter_2019_4}
J.~Richter, J.~Gemmer and R.~Steinigeweg,
\newblock \emph{Impact of eigenstate thermalization on the route to
  equilibrium},
\newblock Phys. Rev. E \textbf{99}(5), 050104(R) (2019),
\newblock \doi{10.1103/physreve.99.050104}.

\bibitem{Dupont_2020}
M.~Dupont and J.~E. Moore,
\newblock \emph{Universal spin dynamics in infinite-temperature one-dimensional
  quantum magnets},
\newblock Phys. Rev. B \textbf{101}(12), 121106(R) (2020),
\newblock \doi{10.1103/physrevb.101.121106}.

\bibitem{Garc_a_Pintos_2017}
L.~P. Garc{\'{\i}}a-Pintos, N.~Linden, A.~S. Malabarba, A.~J. Short and
  A.~Winter,
\newblock \emph{Equilibration time scales of physically relevant observables},
\newblock Phys. Rev. X \textbf{7}(3), 031027 (2017),
\newblock \doi{10.1103/physrevx.7.031027}.

\bibitem{Wilming_2018}
H.~Wilming, T.~R. de~Oliveira, A.~J. Short and J.~Eisert,
\newblock \emph{Equilibration times in closed quantum many-body systems},
\newblock In \emph{Fundamental Theories of Physics}, pp. 435--455. Springer
  International Publishing,
\newblock \doi{10.1007/978-3-319-99046-0_18} (2018).

\bibitem{Alhambra_2020}
{\'{A}}.~M. Alhambra, J.~Riddell and L.~P. Garc{\'{\i}}a-Pintos,
\newblock \emph{Time evolution of correlation functions in quantum many-body
  systems},
\newblock Phys. Rev. Lett. \textbf{124}(11), 110605 (2020),
\newblock \doi{10.1103/physrevlett.124.110605}.

\bibitem{D_Alessio_2016}
L.~D'Alessio, Y.~Kafri, A.~Polkovnikov and M.~Rigol,
\newblock \emph{From quantum chaos and eigenstate thermalization to statistical
  mechanics and thermodynamics},
\newblock Advances in Physics \textbf{65}(3), 239 (2016),
\newblock \doi{10.1080/00018732.2016.1198134}.

\bibitem{Borgonovi_2016}
F.~Borgonovi, F.~Izrailev, L.~Santos and V.~Zelevinsky,
\newblock \emph{Quantum chaos and thermalization in isolated systems of
  interacting particles},
\newblock Phys. Rep. \textbf{626}, 1 (2016),
\newblock \doi{10.1016/j.physrep.2016.02.005}.

\bibitem{Gogolin_2016}
C.~Gogolin and J.~Eisert,
\newblock \emph{Equilibration, thermalisation, and the emergence of statistical
  mechanics in closed quantum systems},
\newblock Rep. Prog. Phys. \textbf{79}(5), 056001 (2016),
\newblock \doi{10.1088/0034-4885/79/5/056001}.

\bibitem{Deutsch_1991}
J.~M. Deutsch,
\newblock \emph{Quantum statistical mechanics in a closed system},
\newblock Phys. Rev. A \textbf{43}(4), 2046 (1991),
\newblock \doi{10.1103/physreva.43.2046}.

\bibitem{Srednicki_1994}
M.~Srednicki,
\newblock \emph{Chaos and quantum thermalization},
\newblock Phys. Rev. E \textbf{50}(2), 888 (1994),
\newblock \doi{10.1103/physreve.50.888}.

\bibitem{Rigol_2008}
M.~Rigol, V.~Dunjko and M.~Olshanii,
\newblock \emph{Thermalization and its mechanism for generic isolated quantum
  systems},
\newblock Nature \textbf{452}(7189), 854 (2008),
\newblock \doi{10.1038/nature06838}.

\bibitem{Santos_2010}
L.~F. Santos and M.~Rigol,
\newblock \emph{Localization and the effects of symmetries in the
  thermalization properties of one-dimensional quantum systems},
\newblock Phys. Rev. E \textbf{82}(3), 031130 (2010),
\newblock \doi{10.1103/physreve.82.031130}.

\bibitem{Steinigeweg_2013}
R.~Steinigeweg, J.~Herbrych and P.~Prelov{\v{s}}ek,
\newblock \emph{Eigenstate thermalization within isolated spin-chain systems},
\newblock Phys. Rev. E \textbf{87}(1), 012118 (2013),
\newblock \doi{10.1103/physreve.87.012118}.

\bibitem{Beugeling_2014}
W.~Beugeling, R.~Moessner and M.~Haque,
\newblock \emph{Finite-size scaling of eigenstate thermalization},
\newblock Phys. Rev. E \textbf{89}(4), 042112 (2014),
\newblock \doi{10.1103/physreve.89.042112}.

\bibitem{Mondaini_2016}
R.~Mondaini, K.~R. Fratus, M.~Srednicki and M.~Rigol,
\newblock \emph{Eigenstate thermalization in the two-dimensional transverse
  field ising model},
\newblock Phys. Rev. E \textbf{93}(3), 032104 (2016),
\newblock \doi{10.1103/physreve.93.032104}.

\bibitem{Mondaini_2017}
R.~Mondaini and M.~Rigol,
\newblock \emph{Eigenstate thermalization in the two-dimensional transverse
  field ising model. {II}. off-diagonal matrix elements of observables},
\newblock Phys. Rev. E \textbf{96}(1), 012157 (2017),
\newblock \doi{10.1103/physreve.96.012157}.

\bibitem{Kim_2014}
H.~Kim, T.~N. Ikeda and D.~A. Huse,
\newblock \emph{Testing whether all eigenstates obey the eigenstate
  thermalization hypothesis},
\newblock Phys. Rev. E \textbf{90}(5), 052105 (2014),
\newblock \doi{10.1103/physreve.90.052105}.

\bibitem{Jansen_2019}
D.~Jansen, J.~Stolpp, L.~Vidmar and F.~Heidrich-Meisner,
\newblock \emph{Eigenstate thermalization and quantum chaos in the Holstein
  polaron model},
\newblock Phys. Rev. B \textbf{99}(15), 155130 (2019),
\newblock \doi{10.1103/physrevb.99.155130}.

\bibitem{Essler_2016}
F.~H.~L. Essler and M.~Fagotti,
\newblock \emph{Quench dynamics and relaxation in isolated integrable quantum
  spin chains},
\newblock J. Stat. Mech: Theory Exp. \textbf{2016}(6), 064002 (2016),
\newblock \doi{10.1088/1742-5468/2016/06/064002}.

\bibitem{Jaynes_1957}
E.~T. Jaynes,
\newblock \emph{Information theory and statistical mechanics},
\newblock Physical Review \textbf{106}(4), 620 (1957),
\newblock \doi{10.1103/physrev.106.620}.

\bibitem{Rigol_2007}
M.~Rigol, V.~Dunjko, V.~Yurovsky and M.~Olshanii,
\newblock \emph{Relaxation in a completely integrable many-body quantum system:
  An ab initio study of the dynamics of the highly excited states of 1d lattice
  hard-core bosons},
\newblock Phys. Rev. Lett. \textbf{98}(5), 050405 (2007),
\newblock \doi{10.1103/physrevlett.98.050405}.

\bibitem{Vidmar_2016}
L.~Vidmar and M.~Rigol,
\newblock \emph{Generalized Gibbs ensemble in integrable lattice models},
\newblock J. Stat. Mech: Theory Exp. \textbf{2016}(6), 064007 (2016),
\newblock \doi{10.1088/1742-5468/2016/06/064007}.

\bibitem{Nandkishore_2015}
R.~Nandkishore and D.~A. Huse,
\newblock \emph{Many-body localization and thermalization in quantum
  statistical mechanics},
\newblock Annu. Rev. Condens. Matter Phys. \textbf{6}(1), 15 (2015),
\newblock \doi{10.1146/annurev-conmatphys-031214-014726}.

\bibitem{Abanin_2019}
D.~A. Abanin, E.~Altman, I.~Bloch and M.~Serbyn,
\newblock \emph{Colloquium: Many-body localization, thermalization, and
  entanglement},
\newblock Rev. Mod. Phys. \textbf{91}(2), 021001 (2019),
\newblock \doi{10.1103/revmodphys.91.021001}.

\bibitem{Shiraishi_2017}
N.~Shiraishi and T.~Mori,
\newblock \emph{Systematic construction of counterexamples to the eigenstate
  thermalization hypothesis},
\newblock Phys. Rev. Lett. \textbf{119}(3), 030601 (2017),
\newblock \doi{10.1103/physrevlett.119.030601}.

\bibitem{Turner_2018}
C.~J. Turner, A.~A. Michailidis, D.~A. Abanin, M.~Serbyn and Z.~Papi{\'{c}},
\newblock \emph{Weak ergodicity breaking from quantum many-body scars},
\newblock Nat. Phys. \textbf{14}(7), 745 (2018),
\newblock \doi{10.1038/s41567-018-0137-5}.

\bibitem{Schecter_2019}
M.~Schecter and T.~Iadecola,
\newblock \emph{Weak ergodicity breaking and quantum many-body scars in spin-1
  {XY} magnets},
\newblock Phys. Rev. Lett. \textbf{123}(14), 147201 (2019),
\newblock \doi{10.1103/physrevlett.123.147201}.

\bibitem{Iadecola_2019}
T.~Iadecola and M.~{\v{Z}}nidari{\v{c}},
\newblock \emph{Exact localized and ballistic eigenstates in disordered chaotic
  spin ladders and the Fermi-Hubbard model},
\newblock Phys. Rev. Lett. \textbf{123}(3), 036403 (2019),
\newblock \doi{10.1103/physrevlett.123.036403}.

\bibitem{Lee_2020}
K.~Lee, R.~Melendrez, A.~Pal and H.~J. Changlani,
\newblock \emph{Exact three-colored quantum scars from geometric frustration}
 \newblock Phys. Rev. B \textbf{101}(24), 241111(R) (2020),
\newblock \doi{10.1103/PhysRevB.101.241111}.

\bibitem{Khemani_2019}
V.~Khemani M.~Hermele and R.~Nandkishore,
\newblock \emph{Local constraints can globally shatter Hilbert space: a new
  route to quantum information protection}
 \newblock Phys. Rev. B \textbf{101}(17), 174204 (2020),
\newblock \doi{10.1103/PhysRevB.101.174204}.

\bibitem{Sala_2020}
P.~Sala, T.~Rakovszky, R.~Verresen, M.~Knap and F.~Pollmann,
\newblock \emph{Ergodicity breaking arising from Hilbert space fragmentation in
  dipole-conserving hamiltonians},
\newblock Phys. Rev. X \textbf{10}(1), 011047 (2020),
\newblock \doi{10.1103/PhysRevX.10.011047}.

\bibitem{Aoki_2014}
H.~Aoki, N.~Tsuji, M.~Eckstein, M.~Kollar, T.~Oka and P.~Werner,
\newblock \emph{Nonequilibrium dynamical mean-field theory and its
  applications},
\newblock Rev. Mod. Phys. \textbf{86}(2), 779 (2014),
\newblock \doi{10.1103/revmodphys.86.779}.

\bibitem{Nauts_1983}
A.~Nauts and R.~E. Wyatt,
\newblock \emph{New approach to many-state quantum dynamics: The
  recursive-residue-generation method},
\newblock Phys. Rev. Lett. \textbf{51}(25), 2238 (1983),
\newblock \doi{10.1103/physrevlett.51.2238}.

\bibitem{Long_2003}
M.~W. Long, P.~Prelov{\v{s}}ek, S.~E. Shawish, J.~Karadamoglou and X.~Zotos,
\newblock \emph{Finite-temperature dynamical correlations using the
  microcanonical ensemble and the Lanczos algorithm},
\newblock Phys. Rev. B \textbf{68}(23), 235106 (2003),
\newblock \doi{10.1103/physrevb.68.235106}.

\bibitem{Heitmann_2020}
T.~Heitmann, J.~Richter, D.~Schubert and R.~Steinigeweg,
\newblock \emph{Selected applications of typicality to real-time dynamics of
  quantum many-body systems}
\newblock Z. Naturforsch. A \textbf{75}, 421 (2020),
\newblock \doi{10.1515/zna-2020-0010}.

\bibitem{Wurtz_2018}
J.~Wurtz, A.~Polkovnikov and D.~Sels,
\newblock \emph{Cluster truncated Wigner approximation in strongly interacting
  systems},
\newblock Ann. Phys. \textbf{395}, 341 (2018),
\newblock \doi{10.1016/j.aop.2018.06.001}.

\bibitem{Schollw_ck_2011}
U.~Schollwöck,
\newblock \emph{The density-matrix renormalization group in the age of matrix
  product states},
\newblock Ann. Phys. \textbf{326}(1), 96 (2011),
\newblock \doi{10.1016/j.aop.2010.09.012}.

\bibitem{Paeckel_2019}
S.~Paeckel, T.~Köhler, A.~Swoboda, S.~R. Manmana, U.~Schollwöck and C.~Hubig,
\newblock \emph{Time-evolution methods for matrix-product states},
\newblock Ann. Phys. \textbf{411}, 167998 (2019),
\newblock \doi{10.1016/j.aop.2019.167998}.

\bibitem{Zaletel_2015}
M.~P. Zaletel, R.~S.~K. Mong, C.~Karrasch, J.~E. Moore and F.~Pollmann,
\newblock \emph{Time-evolving a matrix product state with long-ranged
  interactions},
\newblock Phys. Rev. B \textbf{91}(16), 165112 (2015),
\newblock \doi{10.1103/physrevb.91.165112}.

\bibitem{James_2015}
A.~J.~A. James and R.~M. Konik,
\newblock \emph{Quantum quenches in two spatial dimensions using chain array
  matrix product states},
\newblock Phys. Rev. B \textbf{92}(16), 161111 (2015),
\newblock \doi{10.1103/physrevb.92.161111}.

\bibitem{Hashizume_2018}
T.~Hashizume, I.~P. McCulloch and J.~C. Halimeh,
\newblock \emph{Dynamical phase transitions in the two-dimensional
  transverse-field Ising model} \eprint{https://arXiv.org/abs/1811.09275}.

\bibitem{Hubig_2019}
C.~Hubig and J.~I. Cirac,
\newblock \emph{Time-dependent study of disordered models with infinite
  projected entangled pair states},
\newblock {SciPost} Physics \textbf{6}(3), 031 (2019),
\newblock \doi{10.21468/scipostphys.6.3.031}.

\bibitem{Czarnik_2019}
P.~Czarnik, J.~Dziarmaga and P.~Corboz,
\newblock \emph{Time evolution of an infinite projected entangled pair state:
  An efficient algorithm},
\newblock Phys. Rev. B \textbf{99}(3), 035115 (2019),
\newblock \doi{10.1103/PhysRevB.99.035115}.

\bibitem{Kloss_2020}
B.~Kloss, Y.~Bar Lev and D.~R. Reichman,
\newblock \emph{Studying dynamics in two-dimensional quantum lattices using
  tree tensor network states} \eprint{https://arXiv.org/abs/2003.08944}.

\bibitem{Carleo_2017}
G.~Carleo and M.~Troyer,
\newblock \emph{Solving the quantum many-body problem with artificial neural
  networks},
\newblock Science \textbf{355}(6325), 602 (2017),
\newblock \doi{10.1126/science.aag2302}.

\bibitem{Schmitt_2018}
M.~Schmitt and M.~Heyl,
\newblock \emph{Quantum dynamics in transverse-field Ising models from
  classical networks},
\newblock {SciPost} Physics \textbf{4}(2), 013 (2018),
\newblock \doi{10.21468/SciPostPhys.4.2.013}.

\bibitem{Schmitt_2019}
M.~Schmitt and M.~Heyl,
\newblock \emph{Quantum many-body dynamics in two dimensions with artificial
  neural networks} \eprint{https://arXiv.org/abs/1912.08828}.

\bibitem{Choi_2016}
J.~y.~Choi, S.~Hild, J.~Zeiher, P.~Schauss, A.~Rubio-Abadal, T.~Yefsah,
  V.~Khemani, D.~A. Huse, I.~Bloch and C.~Gross,
\newblock \emph{Exploring the many-body localization transition in two
  dimensions},
\newblock Science \textbf{352}(6293), 1547 (2016),
\newblock \doi{10.1126/science.aaf8834}.

\bibitem{Guardado_Sanchez_2018}
E.~Guardado-Sanchez, P.~T. Brown, D.~Mitra, T.~Devakul, D.~A. Huse,
  P.~Schau{\ss} and W.~S. Bakr,
\newblock \emph{Probing the quench dynamics of antiferromagnetic correlations
  in a 2d quantum Ising spin system},
\newblock Phys. Rev. X \textbf{8}(2), 021069 (2018),
\newblock \doi{10.1103/physrevx.8.021069}.

\bibitem{Lienhard_2018}
V.~Lienhard, S.~de~L{\'{e}}s{\'{e}}leuc, D.~Barredo, T.~Lahaye, A.~Browaeys,
  M.~Schuler, L.-P. Henry and A.~M. Läuchli,
\newblock \emph{Observing the space- and time-dependent growth of correlations
  in dynamically tuned synthetic ising models with antiferromagnetic
  interactions},
\newblock Phys. Rev. X \textbf{8}(2), 021070 (2018),
\newblock \doi{10.1103/physrevx.8.021070}.

\bibitem{Pfeuty_1970}
P.~Pfeuty,
\newblock \emph{The one-dimensional Ising model with a transverse field},
\newblock Ann. Phys. \textbf{57}(1), 79 (1970),
\newblock \doi{10.1016/0003-4916(70)90270-8}.

\bibitem{Bla__2016}
B.~Bla{\ss} and H.~Rieger,
\newblock \emph{Test of quantum thermalization in the two-dimensional
  transverse-field Ising model},
\newblock Sci. Rep. \textbf{6}(1), 38185 (2016),
\newblock \doi{10.1038/srep38185}.

\bibitem{Bl_te_2002}
H.~W.~J. Blöte and Y.~Deng,
\newblock \emph{Cluster Monte Carlo simulation of the transverse Ising model},
\newblock Phys. Rev. E \textbf{66}(6), 066110 (2002),
\newblock \doi{10.1103/physreve.66.066110}.

\bibitem{Rigol_2006}
M.~Rigol, T.~Bryant and R.~R.~P. Singh,
\newblock \emph{Numerical linked-cluster approach to quantum lattice models},
\newblock Phys. Rev. Lett. \textbf{97}(18), 187202 (2006),
\newblock \doi{10.1103/physrevlett.97.187202}.

\bibitem{Rigol_2007_2}
M.~Rigol, T.~Bryant and R.~R.~P. Singh,
\newblock \emph{Numerical linked-cluster algorithms. i. spin systems on square,
  triangular, and kagom{\'{e}} lattices},
\newblock Phys. Rev. E \textbf{75}(6), 061118 (2007),
\newblock \doi{10.1103/physreve.75.061118}.

\bibitem{Ixert_2015}
D.~Ixert, T.~Tischler and K.~P. Schmidt,
\newblock \emph{Nonperturbative linked-cluster expansions for the trimerized
  ground state of the spin-one Kagome Heisenberg model},
\newblock Phys. Rev. B \textbf{92}(17), 174422 (2015),
\newblock \doi{10.1103/physrevb.92.174422}.

\bibitem{Bhattaram_2019}
K.~Bhattaram and E.~Khatami,
\newblock \emph{Lanczos-boosted numerical linked-cluster expansion for quantum
  lattice models},
\newblock Phys. Rev. E \textbf{100}(1), 013305 (2019),
\newblock \doi{10.1103/physreve.100.013305}.

\bibitem{Schaefer_2020}
R.~Schäfer, I.~Hagymási, R.~Moessner and D.~J. Luitz,
\newblock \emph{The pyrochlore s=1/2 Heisenberg antiferromagnet at finite
  temperature} \eprint{https://arXiv.org/abs/2003.04898}.

\bibitem{Kallin_2013}
A.~B. Kallin, K.~Hyatt, R.~R.~P. Singh and R.~G. Melko,
\newblock \emph{Entanglement at a two-dimensional quantum critical point: A
  numerical linked-cluster expansion study},
\newblock Phys. Rev. Lett. \textbf{110}(13), 135702 (2013),
\newblock \doi{10.1103/physrevlett.110.135702}.

\bibitem{Biella_2018}
A.~Biella, J.~Jin, O.~Viyuela, C.~Ciuti, R.~Fazio and D.~Rossini,
\newblock \emph{Linked cluster expansions for open quantum systems on a
  lattice},
\newblock Phys. Rev. B \textbf{97}(3), 035103 (2018),
\newblock \doi{10.1103/physrevb.97.035103}.

\bibitem{Rigol_2014}
M.~Rigol,
\newblock \emph{Quantum quenches in the thermodynamic limit},
\newblock Phys. Rev. Lett. \textbf{112}(17), 170601 (2014),
\newblock \doi{10.1103/physrevlett.112.170601}.

\bibitem{Wouters_2014}
B.~Wouters, J.~D. Nardis, M.~Brockmann, D.~Fioretto, M.~Rigol and J.-S. Caux,
\newblock \emph{Quenching the anisotropic Heisenberg chain: Exact solution and
  generalized Gibbs ensemble predictions},
\newblock Phys. Rev. Lett. \textbf{113}(11), 117202 (2014),
\newblock \doi{10.1103/physrevlett.113.117202}.

\bibitem{White_2017}
I.~G. White, B.~Sundar and K.~R.~A. Hazzard,
\newblock \emph{Quantum dynamics from a numerical linked cluster expansion}
  \eprint{https://arXiv.org/abs/1710.07696}.

\bibitem{Mallayya_2017}
K.~Mallayya and M.~Rigol,
\newblock \emph{Numerical linked cluster expansions for quantum quenches in
  one-dimensional lattices},
\newblock Phys. Rev. E \textbf{95}(3), 033302 (2017),
\newblock \doi{10.1103/physreve.95.033302}.

\bibitem{Mallayya_2018}
K.~Mallayya and M.~Rigol,
\newblock \emph{Quantum quenches and relaxation dynamics in the thermodynamic
  limit},
\newblock Phys. Rev. Lett. \textbf{120}(7), 070603 (2018),
\newblock \doi{10.1103/physrevlett.120.070603}.

\bibitem{Richter_2019}
J.~Richter and R.~Steinigeweg,
\newblock \emph{Combining dynamical quantum typicality and numerical linked
  cluster expansions},
\newblock Phys. Rev. B \textbf{99}(9), 094419 (2019),
\newblock \doi{10.1103/PhysRevB.99.094419}.

\bibitem{Richter_2019_2}
J.~Richter, F.~Jin, L.~Knipschild, H.~De Raedt, K.~Michielsen, J.~Gemmer and
  R.~Steinigeweg,
\newblock \emph{Exponential damping induced by random and realistic
  perturbations} 
\newblock Phys. Rev. E \textbf{101}(6), 062133 (2020),
\newblock \doi{10.1103/PhysRevE.101.062133}.

\bibitem{Tang_2013}
B.~Tang, E.~Khatami and M.~Rigol,
\newblock \emph{A short introduction to numerical linked-cluster expansions},
\newblock Comput. Phys. Commun. \textbf{184}(3), 557 (2013),
\newblock \doi{10.1016/j.cpc.2012.10.008}.

\bibitem{Dusuel_2010}
S.~Dusuel, M.~Kamfor, K.~P. Schmidt, R.~Thomale and J.~Vidal,
\newblock \emph{Bound states in two-dimensional spin systems near the Ising
  limit: A quantum finite-lattice study},
\newblock Physical Review B \textbf{81}(6), 064412 (2010),
\newblock \doi{10.1103/physrevb.81.064412}.

\bibitem{de_Vries_1993}
P.~de~Vries and H.~De Raedt,
\newblock \emph{Solution of the time-dependent Schrödinger equation for
  two-dimensional spin-1/2 Heisenberg systems},
\newblock Physical Review B \textbf{47}(13), 7929 (1993),
\newblock \doi{10.1103/physrevb.47.7929}.

\bibitem{Elsayed_2013}
T.~A. Elsayed and B.~V. Fine,
\newblock \emph{Regression relation for pure quantum states and its
  implications for efficient computing},
\newblock Phys. Rev. Lett. \textbf{110}(7), 070404 (2013),
\newblock \doi{10.1103/PhysRevLett.110.070404}.

\bibitem{Steinigeweg_2014}
R.~Steinigeweg, J.~Gemmer and W.~Brenig,
\newblock \emph{Spin-current autocorrelations from single pure-state
  propagation},
\newblock Phys. Rev. Lett. \textbf{112}(12), 120601 (2014),
\newblock \doi{10.1103/PhysRevLett.112.120601}.

\bibitem{Tal_Ezer_1984}
H.~Tal-Ezer and R.~Kosloff,
\newblock \emph{An accurate and efficient scheme for propagating the time
  dependent Schrödinger equation},
\newblock J. Chem. Phys. \textbf{81}(9), 3967 (1984),
\newblock \doi{10.1063/1.448136}.

\bibitem{Dobrovitski_2003}
V.~V. Dobrovitski and H.~De Raedt,
\newblock \emph{Efficient scheme for numerical simulations of the spin-bath
  decoherence},
\newblock Phys. Rev. E \textbf{67}(5), 056702 (2003),
\newblock \doi{10.1103/physreve.67.056702}.

\bibitem{Wei_e_2006}
A.~Wei{\ss}e, G.~Wellein, A.~Alvermann and H.~Fehske,
\newblock \emph{The kernel polynomial method},
\newblock Rev. Mod. Phys. \textbf{78}(1), 275 (2006),
\newblock \doi{10.1103/revmodphys.78.275}.

\bibitem{Fehske_2009}
H.~Fehske, J.~Schleede, G.~Schubert, G.~Wellein, V.~S. Filinov and A.~R.
  Bishop,
\newblock \emph{Numerical approaches to time evolution of complex quantum
  systems},
\newblock Phys. Lett. A \textbf{373}(25), 2182 (2009),
\newblock \doi{10.1016/j.physleta.2009.04.022}.

\bibitem{Barouch_1970}
E.~Barouch, B.~M. McCoy and M.~Dresden,
\newblock \emph{Statistical mechanics of the XY model. I},
\newblock Phys. Rev. A \textbf{2}(3), 1075 (1970),
\newblock \doi{10.1103/physreva.2.1075}.

\bibitem{Barouch_1971}
E.~Barouch and B.~M. McCoy,
\newblock \emph{Statistical mechanics of the XY model. {II}. spin-correlation
  functions},
\newblock Phys. Rev. A \textbf{3}(2), 786 (1971),
\newblock \doi{10.1103/physreva.3.786}.

\bibitem{Barouch_1971_2}
E.~Barouch and B.~M. McCoy,
\newblock \emph{Statistical mechanics of the XY model. {III}},
\newblock Phys. Rev. A \textbf{3}(6), 2137 (1971),
\newblock \doi{10.1103/physreva.3.2137}.

\bibitem{Igl_i_2000}
F.~Igl{\'{o}}i and H.~Rieger,
\newblock \emph{Long-range correlations in the nonequilibrium quantum
  relaxation of a spin chain},
\newblock Phys. Rev. Lett. \textbf{85}(15), 3233 (2000),
\newblock \doi{10.1103/physrevlett.85.3233}.

\bibitem{Sengupta_2004}
K.~Sengupta, S.~Powell and S.~Sachdev,
\newblock \emph{Quench dynamics across quantum critical points},
\newblock Phys. Rev. A \textbf{69}(5), 053616 (2004),
\newblock \doi{10.1103/physreva.69.053616}.

\bibitem{Rossini_2009}
D.~Rossini, A.~Silva, G.~Mussardo and G.~E. Santoro,
\newblock \emph{Effective thermal dynamics following a quantum quench in a spin
  chain},
\newblock Phys. Rev. Lett. \textbf{102}(12), 127204 (2009),
\newblock \doi{10.1103/physrevlett.102.127204}.

\bibitem{Igl_i_2011}
F.~Igl{\'{o}}i and H.~Rieger,
\newblock \emph{Quantum relaxation after a quench in systems with boundaries},
\newblock Phys. Rev. Lett. \textbf{106}(3), 035701 (2011),
\newblock \doi{10.1103/physrevlett.106.035701}.

\bibitem{Foini_2011}
L.~Foini, L.~F. Cugliandolo and A.~Gambassi,
\newblock \emph{Fluctuation-dissipation relations and critical quenches in the
  transverse field Ising chain},
\newblock Phys. Rev. B \textbf{84}(21), 212404 (2011),
\newblock \doi{10.1103/physrevb.84.212404}.

\bibitem{Calabrese_2011}
P.~Calabrese, F.~H.~L. Essler and M.~Fagotti,
\newblock \emph{Quantum quench in the transverse-field Ising chain},
\newblock Phys. Rev. Lett. \textbf{106}(22), 227203 (2011),
\newblock \doi{10.1103/physrevlett.106.227203}.

\bibitem{Calabrese_2012}
P.~Calabrese, F.~H.~L. Essler and M.~Fagotti,
\newblock \emph{Quantum quench in the transverse field Ising chain: I. time
  evolution of order parameter correlators},
\newblock J. Stat. Mech: Theory Exp. \textbf{2012}(07), P07016 (2012),
\newblock \doi{10.1088/1742-5468/2012/07/p07016}.

\bibitem{Calabrese_2012_2}
P.~Calabrese, F.~H.~L. Essler and M.~Fagotti,
\newblock \emph{Quantum quenches in the transverse field Ising chain: {II}.
  stationary state properties},
\newblock J. Stat. Mech: Theory Exp.
  \textbf{2012}(07), P07022 (2012),
\newblock \doi{10.1088/1742-5468/2012/07/p07022}.

\bibitem{vanVoorden_2020}
B.~van Voorden, J.~Minář and K.~Schoutens,
\newblock \emph{Quantum many-body scars in transverse field Ising ladders and
  beyond} 
\newblock Phys. Rev. B \textbf{101}, 220305(R) (2020),
\newblock \doi{10.1103/PhysRevB.101.220305}.

\bibitem{Hafner_2016}
J.~Hafner, B.~Blass and H.~Rieger,
\newblock \emph{Light cone in the two-dimensional transverse-field Ising model
  in time-dependent mean-field theory},
\newblock {EPL} (Europhysics Letters) \textbf{116}(6), 60002 (2016),
\newblock \doi{10.1209/0295-5075/116/60002}.

\bibitem{De_Nicola_2019}
S.~De Nicola, B.~Doyon and M.~J. Bhaseen,
\newblock \emph{Stochastic approach to non-equilibrium quantum spin systems},
\newblock J. Phys. A: Math. Theor. \textbf{52}(5), 05LT02 (2019),
\newblock \doi{10.1088/1751-8121/aaf9be}.

\bibitem{Bruognolo_2017}
B.~Bruognolo, Z.~Zhu, S.~R. White and E.~M. Stoudenmire,
\newblock \emph{Matrix product state techniques for two-dimensional systems at
  finite temperature} \eprint{https://arXiv.org/abs/1705.05578}.

\bibitem{Gan_2020}
J.~Gan and K.~R.~A. Hazzard,
\newblock \emph{Numerical linked cluster expansions for inhomogeneous systems}
  \eprint{https://arxiv.org/abs/2005.03177}.

\bibitem{Eisert_2010}
J.~Eisert, M.~Cramer and M.~B. Plenio,
\newblock \emph{Colloquium: Area laws for the entanglement entropy},
\newblock Rev. Mod. Phys. \textbf{82}(1), 277 (2010),
\newblock \doi{10.1103/revmodphys.82.277}.

\bibitem{Alba_2009}
V.~Alba, M.~Fagotti and P.~Calabrese,
\newblock \emph{Entanglement entropy of excited states},
\newblock J. Stat. Mech: Theory Exp.
  \textbf{2009}(10), P10020 (2009),
\newblock \doi{10.1088/1742-5468/2009/10/p10020}.

\bibitem{Beugeling_2015}
W.~Beugeling, A.~Andreanov and M.~Haque,
\newblock \emph{Global characteristics of all eigenstates of local many-body
  Hamiltonians: participation ratio and entanglement entropy},
\newblock J. Stat. Mech: Theory Exp.
  \textbf{2015}(2), P02002 (2015),
\newblock \doi{10.1088/1742-5468/2015/02/p02002}.

\bibitem{Reimann_2020}
P.~Reimann, B.~N. Balz, J.~Richter and R.~Steinigeweg,
\newblock \emph{Temporal relaxation of gapped many-body quantum systems},
\newblock Phys. Rev. B \textbf{101}(9), 094302 (2020),
\newblock \doi{10.1103/PhysRevB.101.094302}.

\bibitem{Page_1993}
D.~N. Page,
\newblock \emph{Average entropy of a subsystem},
\newblock Phys. Rev. Lett. \textbf{71}(9), 1291 (1993),
\newblock \doi{10.1103/physrevlett.71.1291}.

\end{thebibliography}

\nolinenumbers

\end{document}